\documentclass[12pt]{article}
\PassOptionsToPackage{dvipsnames}{xcolor}
\pdfoutput=1
\usepackage{amsmath,amssymb,amsthm,bm,graphicx,float,array,multirow,multicol,rotfloat,caption,subcaption,hyperref,cleveref,enumerate,geometry,mathdots,adjustbox,booktabs,parskip,mathtools,tikz,tikz-cd,pdflscape,csquotes,lscape,rotating,empheq,mathrsfs,longtable}
\usepackage[para]{threeparttable}
\usepackage[all]{xy}
\usepackage[normalem]{ulem}
\usepackage[aligntableaux=center]{ytableau}
\usepackage[numbers,sort&compress]{natbib}
\numberwithin{equation}{section}
\numberwithin{figure}{section}
\allowdisplaybreaks
\makeatletter
\usetikzlibrary{arrows,shapes.misc,positioning,decorations.pathmorphing,decorations.markings,decorations.pathreplacing,matrix,patterns,backgrounds}

\definecolor{gold}{RGB}{255,215,0}
\definecolor{purple}{RGB}{160,32,240}
\hypersetup{
    colorlinks=true,
    linkcolor=black,
    citecolor=blue,
    urlcolor=RoyalBlue,
    }
\tikzset{
scale cd/.style={every label/.append style={scale=#1}, cells={nodes={scale=#1}}},
gauge/.style={rounded rectangle, draw=black!100, thick, minimum size=5mm}, 
gaugeD/.style={rounded rectangle, draw=black!100,double,thick,minimum size=5mm},  
empty/.style={rounded rectangle, draw=white!100, thick, minimum size=5mm}, 
flavor/.style={rectangle, draw=black!100, thick, minimum size=5mm},
flavorD/.style={rectangle, draw=black!100, double,thick, minimum size=5mm},
node/.style={circle, thick, draw=black!100,fill=white!100,  minimum size=2mm, inner sep=0pt},
sqnode/.style={rectangle
, thick, draw=black!100,fill=white!100,  minimum size=2mm, inner sep=0pt
},
sonode/.style={circle, thick, draw=black!100,fill=red!100,  minimum size=3mm, inner sep=0pt},
spnode/.style={circle, thick, draw=black!100,fill=blue!100,  minimum size=3mm, inner sep=0pt},
fnode/.style={rectangle, thick, draw=black!100,fill=white!100,  minimum size=3mm, inner sep=0pt},
tnode/.style={rounded rectangle, outer sep=0pt, thick, minimum size=5mm},
brace/.style={decoration={brace, mirror},decorate}
}

\newcommand\notsoscript{\@setfontsize\notsoscript{9}{7}}

\theoremstyle{plain}
\newtheorem*{thm*}{Theorem}

\theoremstyle{definition}

\newtheorem*{defn*}{Definition}

\makeatother

\def\scheme/{our affine scheme} 

\graphicspath{{figures/}}

\begin{document}

\begin{titlepage}
\vspace*{-3cm} 
\begin{flushright}
{\tt DESY-26-059}\\
{\tt CALT-TH 2026-035}
\end{flushright}
\begin{center}
\vspace{1.0cm}

{\LARGE\bfseries The index from a super-point}
\vspace{1.0cm}

{\large
Dhanvarsh Annamalai,$^{1,2}$ 
Hyunook Hur,$^{3}$
Thomas Hussel,$^{1,2}$ \\ Monica Jinwoo Kang,$^{4}$ Craig Lawrie,$^{1,2}$ and Jaewon Song$^{3,5}$\\}
\vspace{.7cm}
{$^1$ II.~Institut für Theoretische Physik, Universität Hamburg, \\ Notkestrasse 9, 22607 Hamburg, Germany}\par
{$^2$ Deutsches Elektronen-Synchrotron DESY,
Notkestrasse 85, 22607 Hamburg, Germany}\par
{$^3$ Department of Physics, Korea Advanced Institute of Science and Technology\\
Daejeon 34141, Republic of Korea}\par
{$^4$ Mitchell Institute for Fundamental Physics and Astronomy, Texas A\&M University,\\
College Station, TX 77843, U.S.A.}\par
{$^5$ Walter Burke Institute for Theoretical Physics \& Leinweber Forum for Theoretical Physics\\ California Institute of Technology, 
Pasadena, CA 91125, USA}\par

\vspace{.3cm}

\scalebox{.72}{\tt dhanvarsh.annamalai@studium.uni-hamburg.de, hyunookh@kaist.ac.kr, thomas.hussel@studium.uni-hamburg.de,}
\scalebox{.72}{\tt monicak@tamu.edu, craig.lawrie1729@gmail.com, jaewon.song@kaist.ac.kr}\par

\vspace{0.7cm}
\textbf{Abstract}
\end{center}
{
We extend the algebro-geometric construction of filtered graded rings that encode both the Higgs branch and the Macdonald index of 4d $\mathcal{N}=2$ superconformal field theories to theories whose Schur sector contains fermionic generators. The rings are then $\mathbb{Z}_2$-graded-commutative, and the index is reproduced by the Hilbert series of a generalized arc space, whose defining ideal may contain relations among the generators and their derivatives. We study the free $\mathcal{N}=2$ vector multiplet, $\mathcal{N}=4$ super Yang--Mills with gauge algebras $\mathfrak{su}(2)$ and $\mathfrak{su}(3)$, the rank-one $\mathcal{N}=3$ SCFTs, and the $\widehat{E}_6(SU(2))$ SCFT. For the rank-one $\mathcal{N}=3$ SCFTs, the ideal is the sum of the Jacobian ideals of a single superpotential and of its two supersymmetry variations. For $\mathcal{N}=4$ super Yang--Mills, the ideal is similarly the sum of the Jacobian ideals of the components of a single $\mathcal{N}=4$ superfield, which is a singlet of the $\mathfrak{su}(2)$ flavor symmetry for gauge algebra $\mathfrak{su}(2)$, and a triplet for $\mathfrak{su}(3)$. The associated VOA for the $\widehat{E}_6(SU(2))$ theory is given by the Feigin--Feigin--Tipunin $\mathcal{A}(6)$ algebra. The arc space of Zhu's $C_2$-algebra of $\mathcal{A}(6)$ does not reproduce the vacuum character, and therefore $\mathcal{A}(6)$ is not classically free. Supplementing Zhu's $C_2$-algebra by three relations in the first jet ring, the generalized arc space reproduces the Macdonald index of $\widehat{E}_6(SU(2))$ up to, at least, order $q^{51}$.
}
\vfill 
\end{titlepage}


\tableofcontents
\newpage

\section{Introduction}
\label{Intro}

Local operators and their operator product expansion (OPE) lie at the heart of conformal field theory: the spectrum of local operators, together with the OPE coefficients, determines all correlation functions of local operators \cite{Ferrara:1973yt,Polyakov:1974gs}. The consistency of the OPE, as exploited by the conformal bootstrap, leads to constraints which hold uniformly across theories, such as bounds on operator dimensions \cite{Rattazzi:2008pe,Poland:2018epd}. Only in rare cases, however, has it been shown to single out an individual theory. In two dimensions, the Virasoro minimal models are characterized by the decoupling of null states, which constrains them so strongly that they can be solved exactly \cite{Belavin:1984vu}; in three dimensions, the critical Ising and $O(N)$ models have been isolated by the numerical bootstrap \cite{El-Showk:2012cjh,Kos:2014bka,Kos:2016ysd,Chester:2019ifh}. The lesson of two dimensions is that the operators which are absent from an OPE can carry as much information as those which are present. This suggests an algebraic strategy for pinpointing a theory: postulate a set of generating operators, determine the relations among them which are implied by the decoupling of operators in their OPEs, and reconstruct the operator content of the theory from these generators and relations.

Supersymmetric conformal field theories provide an ideal setting in which to make this strategy manifest, since their protected operators have OPEs which are rigid enough to be studied exactly, even at strong coupling. Indeed, protected quantities, such as central charges and superconformal indices, have been used to map out large landscapes of strongly-coupled supersymmetric fixed points, together with the dualities and renormalization group flows among them \cite{Kang:2021ccs,Kang:2022vab,Kang:2023dsa,Kang:2024elv,Kang:2024inx}. Four-dimensional $\mathcal{N}=2$ superconformal field theories (SCFTs), in particular, form a remarkably rich landscape, much of which consists of strongly-coupled theories without a known Lagrangian description, which arise, for example, at special points on the Coulomb branches of gauge theories, from compactifications of 6d SCFTs, from string theory, and by gauging the flavor symmetries of other such theories \cite{Argyres:1995jj,Argyres:1995xn,Minahan:1996fg,Minahan:1996cj,Gaiotto:2009we,Cecotti:2012jx,Cecotti:2013lda,Ohmori:2015pua,Ohmori:2015pia,Baume:2021qho,Kang:2021lic,Distler:2022yse,Distler:2022kjb,Couzens:2023kyf,Giacomelli:2024ycb,Giacomelli:2025zqn}. For these theories, the protected sectors are highly structured: they contain infinitely many operators, and their OPEs are subject to strong selection rules, so that operators which would generically appear in an OPE are often absent \cite{Liendo:2015ofa,Agarwal:2018zqi}. Thus, they provide a natural testing ground for the algebraic strategy outlined above.

Two protected structures of 4d $\mathcal{N}=2$ SCFTs play a central role in this paper. The first is the Higgs branch chiral ring, which describes the geometry of the Higgs branch, $\mathcal{M}_\mathcal{H}$, of the moduli space of vacua. The second is the Schur sector, which consists of the $1/4$-BPS operators counted by the Schur and Macdonald limits of the superconformal index \cite{Gadde:2011uv}; via the SCFT/VOA correspondence, these operators are organized into a two-dimensional vertex operator algebra (VOA), also called the chiral algebra \cite{Beem:2013sza}. The two structures are intimately related: the Higgs branch chiral ring operators belong to the Schur sector, and the Higgs branch is conjecturally recovered from the VOA as its associated variety \cite{Beem:2017ooy,arakawa2015associated}. They nevertheless capture very different amounts of information. The Higgs branch can be a single point even when the Schur sector is highly non-trivial, as happens for many Argyres--Douglas theories, and even the relation between the Higgs branch and the Hall--Littlewood limit of the index is a subtle one \cite{Kang:2022zsl}. It is therefore natural to ask whether there exists a single, simple object which encodes both the Higgs branch and the Schur sector, and from which each of them can be read off.

In a series of papers \cite{Kang:2025zub,Kang:2026nge,Kang:2026ona,Kang:2026ney,Kang:2026jnv}, we have proposed such an object: a ring, $R$, that encodes both the Higgs branch and the Macdonald index, extending the observation of \cite{Bhargava:2023hsc} for the $(A_1, A_{2n})$ Argyres--Douglas theory to a much larger class of 4d $\mathcal{N}=2$ SCFTs.\footnote{See also \cite{Andrews:2025krn} for closely related constructions for the $(A_1, D_{2n+1})$ Argyres--Douglas theories.} The generators of $R$ correspond to certain protected operators, and the relations correspond to the decoupling of operators in their OPEs \cite{Liendo:2015ofa,Agarwal:2018zqi, Song:2021dhu}. For example, the stress tensor multiplet of the $(A_1, A_{2n})$ Argyres--Douglas theory obeys a relation of the schematic form $T^{n+1} \sim 0$, which leads us to the ring $R = \mathbb{C}[x] \, / \, (x^{n+1})$. Via the SCFT/VOA correspondence, this relation is the image of the vacuum null state of the $(2, 2n+3)$ Virasoro minimal model, which is the VOA of this theory \cite{Beem:2013sza,Cordova:2015nma}, so that the two-dimensional decoupling discussed above becomes a relation in $R$. The Higgs branch is recovered as the reduced spectrum of $R$; for $(A_1, A_{2n})$ this is a single point, but the nilpotent elements $x, \cdots, x^n$ retain information about the protected operators which is lost upon passing to the Higgs branch. To also account for derivatives of operators, we replace each generator of $R$ by a formal power series, $x(t) = \sum_{j \geq 0} x^{(j)} t^j$, where $x^{(j)}$ corresponds to $\partial^j x / j!$, and impose every coefficient of each relation in $t$; the resulting ring, $J_\infty(R)$, is the coordinate ring of the arc space of $\operatorname{Spec}(R)$.\footnote{Geometrically, the arc space of a scheme, $X$, parametrizes formal arcs in $X$, that is, maps from the formal disc into $X$. Truncating the power series at order $t^n$ gives the $n$th jet scheme; for smooth $X$, the first jet scheme is the tangent bundle of $X$.} Counting the elements of $J_\infty(R)$ then reproduces the index.

Let us state this proposal more precisely. The Macdonald index \cite{Gadde:2011uv} counts the $1/4$-BPS Schur operators of the SCFT, weighted by their fermion number and by two fugacities, $q$ and~$T$, and is defined as
\begin{equation}
    I_\text{Mac}(q, T) = \operatorname{Tr} \, (-1)^F q^{\Delta - R} T^{R - r} \,,
\end{equation}
where the trace is taken over operators satisfying $\Delta - 2j_1 - 2R - r = 0$ and $j_1 - j_2 + r = 0$. Here, as usual, $\Delta$ is the conformal dimension, $(j_1, j_2)$ are the Lorentz spins, and $R$ and $r$ are the charges under the $\mathfrak{su}(2)_R$ and $\mathfrak{u}(1)_r$ R-symmetries, respectively. We equip the ring $R$ with a $q$-grading together with a compatible increasing, multiplicative $T$-filtration, and refer to $R$ as a filtered graded ring. The proposal is that
\begin{equation}\label{eqn:bigeqn}
    I_\text{Mac}(q, T) = \operatorname{HS}_{p,q,T}(\operatorname{gr}(J_\infty(R))) \big|_{p \rightarrow q} \,.
\end{equation}
The Hilbert series $\operatorname{HS}_{p,q,T}$ keeps track of the jet-degree, that is, the number of derivatives, $j$, of a coefficient $x^{(j)}$, via $p$, and of the bigrading of the associated graded ring, inherited from the filtered graded structure of $R$, via $q$ and $T$. Setting $p = q$ reflects that each derivative increases the conformal dimension by one, and we abbreviate the resulting specialization as $\operatorname{HS}_{q,q,T}$. If the SCFT has flavor symmetries, these can be incorporated as additional gradings on $R$, which are then reflected in the Hilbert series. Furthermore, it is proposed that
\begin{equation}
    \operatorname{Spec}(R)_{\operatorname{red}} = \mathcal{M}_\mathcal{H} \,,
\end{equation}
where $\mathcal{M}_\mathcal{H}$ is the Higgs branch of the 4d SCFT.\footnote{As we discuss anon, this is identical to the proposal of \cite{Beem:2017ooy} which identifies the associated variety \cite{arakawa2015associated} of the VOA as the Higgs branch, once $R$ is identified with Zhu's $C_2$-algebra.} Thus, $R$ can be thought of as capturing a ``scheme-ification'' of the Higgs branch. The Schur index follows by taking $T \rightarrow 1$ in equation \eqref{eqn:bigeqn}; equivalently, we can forget about the $T$-filtration on $R$, and only remember the $q$-grading. We refer to this graded ring as $R_S$, and the proposal for the Schur index is
\begin{equation}\label{eqn:schureqn}
    I_\text{Schur}(q) = \operatorname{HS}_{q,q}(J_\infty(R_S)) \,.
\end{equation}

Determining the decoupling relations directly is often difficult, and so we have developed complementary approaches to finding $R$. For the $(A_{k-1}, A_{N-1})$ Argyres--Douglas theories, which have a trivial Higgs branch, the defining ideal of $R$ is the Jacobian ideal of a single polynomial, the potential, which is uniquely fixed by requiring that $R$ is a complete intersection and that the first jet scheme has a particular minimal syzygy; we refer to the latter condition as ``syzygy maximization'' \cite{Kang:2026nge}. The same structure persists for the $\mathcal{D}_p(SU(N))$ Argyres--Douglas theories, which have non-trivial Higgs branches, and Higgs branch renormalization group flows are realized by restricting the potential to a Slodowy slice \cite{Kang:2026ona}. In \cite{Kang:2026ney}, we gave a uniform construction of the potential for all $\mathcal{D}_p(G)$ theories, and we proposed that, for these Argyres--Douglas theories, the $T$-filtration is intrinsic to the ring: it is the filtration by powers of the nilradical. The critical schemes of these potentials, their jet syzygies, and their relation to twisted periods are studied from a mathematical perspective in \cite{Kang:2026jnv}. Thus, within this proposal, the Higgs branch, the filtered graded ring, and hence the Macdonald index of each of these theories all follow from a single polynomial.

In all of these works, the ring $R$ is commutative, and the generators of $R$ correspond to bosonic operators. However, many 4d $\mathcal{N}=2$ SCFTs have fermionic operators in their Schur sector which cannot be written in terms of other operators. The simplest example is the free $\mathcal{N}=2$ vector multiplet, whose Schur sector contains the gaugini, and whose Macdonald index is \cite{Gadde:2011uv}
\begin{equation}
    I_\text{Mac}^{\mathcal{N}=2\text{ vector}}(q, T) = \operatorname{PE} \left[ \frac{-q - qT}{1-q} \right] = 1 - q(1 + T) - q^2 + \cdots \,.
\end{equation}
Since fermionic operators contribute to the index with a minus sign, the expansion of this index has negative coefficients, and thus it cannot be the Hilbert series of the arc space of any commutative ring, whose coefficients are dimensions of vector spaces and are thus non-negative. The free vector multiplet is far from the only theory with fermionic strong generators, that is, fermionic generators of the VOA which cannot be written as normally ordered products of other operators and their derivatives. Whenever the 4d theory has enhanced $\mathcal{N}=3$ or $\mathcal{N}=4$ supersymmetry, the additional supercurrents give rise to fermionic currents of conformal dimension $3/2$ in the VOA, which, together with the stress tensor and additional bosonic currents, generate the $\mathcal{N}=2$ or the small $\mathcal{N}=4$ superconformal algebra, respectively \cite{Beem:2013sza,Nishinaka:2016hbw}. Fermionic strong generators also arise in interacting theories with only $\mathcal{N}=2$ supersymmetry, as we will see for the $\widehat{E}_6(SU(2))$ SCFT. These examples force us to go beyond commutative rings.

In this paper, we extend the proposal to $\mathbb{Z}_2$-graded-commutative rings,\footnote{That is, rings graded by the fermion number, $|a| \in \mathbb{Z}_2$, in which $ab = (-1)^{|a||b|} ba$; in particular, every fermionic element squares to zero.} which have both bosonic and fermionic generators, and in which the fermionic elements are counted with a sign. Geometrically, $\operatorname{Spec}(R)$ then becomes an affine superscheme. For the free vector multiplet, the ring is simply the exterior algebra on two fermionic generators, $R = \Lambda[y, z]$, one for each gaugino, and its spectrum is the super-point $\mathbb{C}^{0|2}$: a single point equipped with two fermionic directions. Its arc space is the free $\mathbb{Z}_2$-graded-commutative ring on the coefficients $y^{(j)}$ and $z^{(j)}$, which, as a graded vector space, is the vacuum module of the small $bc$-system, and its Hilbert series reproduces the Macdonald index exactly. Continuing the terminology of \cite{Kang:2025zub,Kang:2026ona}, the index of the free vector multiplet thus arises from a super-point. More generally, since the fermionic generators are nilpotent, they are invisible to the reduced spectrum, and hence to the Higgs branch, but they are indispensable for the index.

Let us describe the general framework. Let $S$ be a $\mathbb{Z}_2$-graded-commutative polynomial ring over $\mathbb{C}$, with $m$ bosonic and $r$ fermionic generators,
\begin{equation}
    S = \mathbb{C}[x_1, \cdots, x_m] \otimes \Lambda[y_1, \cdots, y_r] \,,
\end{equation}
where $\Lambda$ denotes the exterior algebra; $S$ is the coordinate ring of the affine superspace $\mathbb{C}^{m|r}$. We equip $S$ with a bigrading, specified via the degree assignments
\begin{equation}
    \operatorname{deg}(x_i) = (\alpha_i, \beta_i) \,, \quad \operatorname{deg}(y_a) = (\gamma_a, \kappa_a) \,.
\end{equation}
As in the commutative case, we consider quotients $R = S \, / \, I$ by ideals generated by polynomials which are homogeneous in the first grading, but not necessarily in the second, so that $R$ is a filtered graded ring. The proposal that $\operatorname{Spec}(R)_\text{red}$ is the Higgs branch carries over unchanged. The $T$-degrees of the generators are fixed by the 4d charges of the corresponding operators, and they carry information that is not determined by the VOA alone: for example, the four supercurrents of $\mathcal{N}=4$ super Yang--Mills with gauge algebra $\mathfrak{su}(2)$, which we study in Section \ref{sec:N4}, all have conformal dimension $3/2$, but two have $T$-degree $1/2$ and two have $T$-degree $3/2$. For the rank-one $\mathcal{N}=3$ SCFTs, which include $\mathcal{N}=4$ super Yang--Mills with gauge algebra $\mathfrak{su}(2)$ as the case $p = 2$, and for $\mathcal{N}=4$ super Yang--Mills with gauge algebra $\mathfrak{su}(3)$, these $T$-degrees have a more intrinsic origin: the two supersymmetry variations, $\delta_\pm$, which relate the bosonic and fermionic generators, raise the conformal dimension by $1/2$ and shift the $T$-degree by $\pm 1/2$.

It remains an open problem to determine how (and if) the full filtered graded ring, $R$, can be derived from the vertex operator algebra. We expect that the $T$-filtration on $R$ should be related to the filtration on the VOA associated to the four-dimensional R-symmetry charges \cite{Song:2016yfd,Beem:2019tfp,ArabiArdehali:2025fad,Beem:2026lkq}; however, there is not a known inherently 2d way to determine this filtration. For the $\mathcal{D}_p(G)$ Argyres--Douglas theories, the nilradical filtration of \cite{Kang:2026ney} provides an intrinsic candidate, but we do not expect it to extend to general 4d $\mathcal{N}=2$ SCFTs, and indeed it does not extend directly to the rings studied in this paper, as we discuss in Section \ref{sec:disc}.

The ring $R$ is closely related to a ring that can be constructed directly from the VOA of the SCFT \cite{Beem:2013sza}, which we denote by $\mathbb{V}$. Associated to $\mathbb{V}$ is a commutative ring,\footnote{We are interested in cases where $\mathbb{V}$ is a vertex operator superalgebra, in which case $R_\mathbb{V}$ is a $\mathbb{Z}_2$-graded-commutative ring. The extension to vertex operator superalgebras is discussed in detail in \cite{MR4340951}.} $R_\mathbb{V}$, known as Zhu's commutative algebra (or Zhu's $C_2$-algebra) \cite{MR1317233}, which is defined as follows:\footnote{See \cite{AM} for standard definitions and notations in the field of VOAs.}
\begin{equation}
    C_2(\mathbb{V}) = \operatorname{Span}\left( a_{(-h_a-1)}b \, | \, a, b \in \mathbb{V} \right) \,, \qquad R_\mathbb{V} = \mathbb{V} \, / \, C_2(\mathbb{V}) \,.
\end{equation}
Here $h_a$ is the conformal dimension of $a$, and we use the standard, physics mode convention which is $Y(a, z) = \sum_n a_{(n)} z^{-n-h_a}$. The grading on $\mathbb{V}$ coming from the conformal dimension descends to a grading on $R_\mathbb{V}$; we leave this structure implicit whenever we write $R_\mathbb{V}$. More generally, $R_{\mathbb{V}}$ forms part of a larger structure, related to the canonical decreasing filtration on $\mathbb{V}$ known as the Li filtration \cite{MR2172688}:
\begin{equation}
    F^k(\mathbb{V}) = \operatorname{Span}\left( a_{(-n_1-h_{a_1})}^1 \, \cdots \, a_{(-n_r-h_{a_r})}^r \right) \,,
\end{equation}
where $a^j$ are elements of $\mathbb{V}$ and the $n_j$ are non-negative integers such that $\sum_j n_j \geq k$. The associated graded algebra,
\begin{equation}
    \operatorname{gr}_{\operatorname{Li}}(\mathbb{V}) = \bigoplus_k F^k(\mathbb{V}) \, / \, F^{k+1}(\mathbb{V}) \,,
\end{equation}
is generated by $R_{\mathbb{V}}$ as a differential algebra \cite{MR2172688}. Since the Li filtration preserves the conformal dimension, the supercharacter of $\mathbb{V}$, which coincides with the Schur index up to an overall factor of $q^{-c/24}$, can be written as a Hilbert series:
\begin{equation}
    \operatorname{sch}_{\mathbb{V}}(q) = q^{-c/24} \, \operatorname{HS}_{q,-1}(\operatorname{gr}_{\operatorname{Li}}(\mathbb{V})) \,,
\end{equation}
where the second fugacity, here set to $-1$, keeps track of the fermion number. Since $\operatorname{gr}_{\operatorname{Li}}(\mathbb{V})$ is generated by $R_{\mathbb{V}}$, there exists a surjective homomorphism of differential algebras
\begin{equation}\label{eqn:CF}
    \Phi :\, J_\infty(R_{\mathbb{V}}) \,\rightarrow\, \,\operatorname{gr}_{\operatorname{Li}}(\mathbb{V}) \,.
\end{equation}
When $\Phi$ is an isomorphism, the VOA is referred to as ``classically free'' \cite{MR4321999,MR4779563}, and then taking $R_S = R_{\mathbb{V}}$ realizes the proposal in equation \eqref{eqn:schureqn}. Classical freeness has been established in a number of cases, see, for example, \cite{MR4779563,MR4668791,MR4340951,MR4256406,MR4806362,kanade2026classicalfreeness}, including, recently, for the affine $\mathfrak{sl}_2$ VOAs at boundary admissible levels \cite{arakawa2026clfrsl2}, which arise from the $\mathcal{D}_p(SU(2))$ theories with $p$ odd; it is also known to fail, for example for the Ising model \cite{MR4589087}. It has been conjectured \cite{RastelliTalk1,RastelliTalk2} that all VOAs arising from 4d $\mathcal{N}=2$ SCFTs via the SCFT/VOA correspondence are classically free, and in \cite{Kang:2026ona,Kang:2026ney} we conjectured that $R_S \cong R_\mathbb{V}$ and that $\mathbb{V}$ is classically free for the Argyres--Douglas theories considered there; the former has since been proved in \cite{Arakawa:2026wwu}. 

In this paper, we provide a counterexample to the conjecture of \cite{RastelliTalk1,RastelliTalk2}: the $\mathcal{A}(6)$ VOA associated to $\widehat{E}_6(SU(2))$ is not classically free; to the best of our knowledge, this is the first example of a VOA arising from a 4d $\mathcal{N}=2$ SCFT via the SCFT/VOA correspondence which is not classically free. The $\widehat{E}_6(SU(2))$ SCFT is obtained by gauging the diagonal $\mathfrak{su}(2)$ flavor symmetry of three copies of the $\mathcal{D}_3(SU(2))$ Argyres--Douglas theory,\footnote{It is the simplest example of a large class of 4d $\mathcal{N}=2$ SCFTs, $\widehat{\Gamma}(G)$, obtained by such gaugings of Argyres--Douglas theories, many of which, including $\widehat{E}_6(SU(2))$ itself, have $a = c$ \cite{Kang:2021lic,Kang:2022vab}; we return to these theories in Section \ref{sec:disc}.} and the $\mathcal{A}(6)$ VOA is strongly generated by the stress tensor, $T$, together with two fermionic Virasoro primaries, $\Psi$ and $\widetilde{\Psi}$, of conformal dimension four.

The failure of classical freeness has a transparent origin. In the free-field realization of $\mathcal{A}(6)$ in terms of a chiral boson, $\varphi$, the generator $\Psi$ is the exponential $e^{-\sqrt{3}\varphi}$, whose OPE with itself vanishes to third order, and, consequently, the normally ordered product ${:}\partial\Psi\Psi{:}$ vanishes identically. In the arc space of $R_\mathbb{V}$, however, the corresponding element, $y^{(0)} y^{(1)}$, where $y$ is the coordinate corresponding to $\Psi$, is non-zero: as for any fermion, $(y^{(0)})^2 = 0$, but nothing forces $y^{(0)} y^{(1)}$ to vanish. As a result, the Hilbert series of $J_\infty(R_\mathbb{V})$ differs from the vacuum character, starting at conformal dimension nine. Strikingly, $\mathcal{A}(6)$ is related to the small $\mathcal{N}=4$ superconformal algebra with $c = -9$, the VOA of $\mathcal{N}=4$ super Yang--Mills with gauge algebra $\mathfrak{su}(2)$, by an exact graded vector space isomorphism \cite{Buican:2020moo}; in contrast to $\mathcal{A}(6)$, our results indicate that the latter is classically free.

To account for such relations, we allow the ideal to contain relations among the generators and their first derivatives, that is, to live in the first jet ring. Let $I_1$ be an ideal of the first jet ring, $S_1 = J_1(S)$, generated by polynomials which are homogeneous in the jet-degree and in the $q$-grading, but, as before, not necessarily in the $T$-grading, so that $S_1 / I_1$ is typically filtered graded rather than bigraded. We propose that the Macdonald index is then reproduced as
\begin{equation}\label{eqn:biggereqn}
    I_\text{Mac}(q, T) = \operatorname{HS}_{q,q,T,-1}(\operatorname{gr}(J_{\infty,1}(S, I_1))) \,,
\end{equation}
where $J_{\infty,1}(S, I_1)$ is the generalized arc space defined in Section \ref{sec:arcs}, which is obtained by imposing the relations in $I_1$ together with all of their derivatives. The Hilbert series $\operatorname{HS}_{p,q,T,\tau}$ includes a fugacity, $\tau$, for the $\mathbb{Z}_2$-grading, and the subscripts in equation \eqref{eqn:biggereqn} indicate that we set $p = q$ and $\tau = -1$. When $r = 0$ and $I_1 = D^{\leq 1}I_0$, the ideal generated by the elements of an ideal $I_0$ of $S$ together with their first derivatives, we recover the proposal in equation \eqref{eqn:bigeqn}. For $\mathcal{A}(6)$, three relations in the first jet ring, the first of which is the vanishing of ${:}\partial\Psi\Psi{:}$, suffice to reproduce the Macdonald index of $\widehat{E}_6(SU(2))$, and thus also the vacuum character, up to and including order $q^{51}$. For every SCFT that we study in this paper, relations at the level of the first jet ring suffice; whether relations involving higher derivatives are necessary for other 4d $\mathcal{N}=2$ SCFTs is an open question. Since there is no analogue of Hilbert's basis theorem for differential ideals \cite{Ritt,Kolchin}, it is not even a priori clear that relations of some finite order suffice; nevertheless, we conjecture in Section \ref{sec:arcs} that this is the case for every VOA arising from a 4d $\mathcal{N}=2$ SCFT, and we return to these questions in Section \ref{sec:disc}.

A second theme of this paper is that, for some of the theories that we study, the ideal is again built from the derivatives of a potential, as for the Argyres--Douglas theories in \cite{Kang:2026nge,Kang:2026ona,Kang:2026ney,Kang:2026jnv}, now with a $\mathbb{Z}_2$-graded twist. For the rank-one $\mathcal{N}=3$ SCFTs, labelled by $\mathbb{Z}_p$ with $p = 2, 3, 4, 6$, we find a single superpotential, $P_p$, such that the defining ideal is
\begin{equation}
    I_p = \operatorname{Jac}(P_p) + \operatorname{Jac}(\delta_+ P_p) + \operatorname{Jac}(\delta_- P_p) \,,
\end{equation}
where $\operatorname{Jac}(f)$ is the Jacobian ideal of $f$, generated by its derivatives with respect to all of the bosonic and fermionic coordinates. This compact formula reproduces the full ideal obtained from the free-field realization of the VOA, it recovers the Higgs branch, $\mathbb{C}^2/\mathbb{Z}_p$, as the reduced spectrum, and, for $p = 2$, it accounts for both the null relations and the Sugawara relation of $\mathcal{N}=4$ super Yang--Mills with gauge algebra $\mathfrak{su}(2)$. For $\mathcal{N}=4$ super Yang--Mills with gauge algebra $\mathfrak{su}(3)$, the same construction requires a triplet of superpotentials, which transforms in the adjoint representation of the $\mathfrak{su}(2)$ flavor symmetry, and no single invariant superpotential suffices. The difference between the two gauge algebras becomes transparent in $\mathcal{N}=4$ superspace: in both cases, the ideal is the sum of the Jacobian ideals of all of the components of a single superfield, which has spin zero for $\mathfrak{su}(2)$ and spin one for $\mathfrak{su}(3)$. We stress that this is a feature of particular families of theories, rather than of the general proposal: in general, we do not expect $R_S$ to be the quotient by a Jacobian ideal.

\paragraph{Summary of results.} The examples that we study, and the main results, are as follows; the orders to which we have checked each result are listed in Table \ref{tbl:summary}.
\begin{itemize}
    \item \textbf{The free $\mathcal{N}=2$ vector multiplet.} The ring is the exterior algebra on two generators, whose spectrum is the super-point $\mathbb{C}^{0|2}$, and the Hilbert series of its arc space reproduces the Macdonald index exactly.
    \item \textbf{$\mathcal{N}=4$ super Yang--Mills.} For gauge algebra $\mathfrak{su}(2)$, the ring is Zhu's $C_2$-algebra of the $c=-9$ small $\mathcal{N}=4$ superconformal algebra, supplemented by a generator for the stress tensor; for $\mathfrak{su}(3)$, the generators form two $\mathcal{N}=4$ multiplets with twenty components in total. In both cases, the ideal is the sum of the Jacobian ideals of the components of a single $\mathcal{N}=4$ superfield, and the Hilbert series of the arc space reproduces the flavor-refined Macdonald index.
    \item \textbf{Rank-one $\mathcal{N}=3$ SCFTs.} For $p = 3, 4, 6$, the ring is the quotient of $S$ by $I_p$, and it coincides with the ring obtained from the free-field realization of the VOA. We compare its Hilbert series with the Macdonald index computed using the code of \cite{Agarwal:2021oyl}, and, for $p = 3$, with the closed-form Schur index of \cite{Guo:2026lsl}, and we present the higher-order terms, to $q^{22}$, $q^{23}$, and $q^{24}$ respectively, of the Macdonald index as a prediction.
    \item \textbf{The $\mathcal{A}(6)$ VOA and the $\widehat{E}_6(SU(2))$ SCFT.} The arc space of $R_\mathbb{V}$ fails to reproduce the vacuum character at conformal dimension nine, and three relations in the first jet ring restore the agreement with the Macdonald index, which we compute by gauging three copies of $\mathcal{D}_3(SU(2))$.
\end{itemize}

\begin{table}[H]
    \centering
    \renewcommand{\arraystretch}{1.3}
    \begin{tabular}{lcccc}
        \toprule
        4d SCFT & $c$ & $\mathbb{C}^{m|r}$ & $k$ & Agreement up to \\
        \midrule
        Free $\mathcal{N}=2$ vector & $-2$ & $\mathbb{C}^{0|2}$ & $0$ & all orders \\
        $\mathcal{N}=4$ SYM, $\mathfrak{su}(2)$ & $-9$ & $\mathbb{C}^{3|4}$ & $0$ & $q^{21}$ \\
        $\mathcal{N}=4$ SYM, $\mathfrak{su}(3)$ & $-24$ & $\mathbb{C}^{8|10}$ & $0$ & $q^{12}$ \\
        Rank-one $\mathcal{N}=3$ ($p = 3, 4, 6$) & $-3(2p-1)$ & $\mathbb{C}^{4|4}$ & $0$ & $q^{10}$ \\
        $\widehat{E}_6(SU(2))$ & $-24$ & $\mathbb{C}^{1|2}$ & $1$ & $q^{51}$ \\
        \bottomrule
    \end{tabular}
    \caption{The 4d $\mathcal{N}=2$ SCFTs studied in this paper. For each theory, we list the central charge, $c$, of the associated VOA, the affine superspace $\mathbb{C}^{m|r}$ associated to the strong generators of the VOA, with $m$ bosonic and $r$ fermionic strong generators, the order, $k$, of the relations in the ideal $I_k$ of $S_k$, and the order in $q$ up to and including which the Hilbert series of $\operatorname{gr}(J_{\infty,k}(S, I_k))$ agrees with the Macdonald index (refined by the $\mathfrak{su}(2)$ flavor symmetry for $\mathcal{N}=4$ super Yang--Mills). When $k = 0$, the generalized arc space is the arc space of $S \, / \, I_0$.
    }
    \label{tbl:summary}
\end{table}

\paragraph{Organization.} The organization of this paper is as follows. In Section \ref{sec:arcs}, we review the arc space and jet scheme constructions, and introduce the generalized arc spaces $J_{\infty,k}(S, I_k)$, including for $\mathbb{Z}_2$-graded-commutative rings. In Section \ref{sec:vec}, we consider the free $\mathcal{N}=2$ vector multiplet. In Section \ref{sec:N4}, we study $\mathcal{N}=4$ super Yang--Mills with gauge algebras $\mathfrak{su}(2)$ and $\mathfrak{su}(3)$, and we give a formulation of the ideals in terms of $\mathcal{N}=4$ superfields. In Section \ref{sec:N3}, we study the rank-one $\mathcal{N}=3$ SCFTs and their superpotentials, and show that the $p = 2$ case reproduces the ring of Section \ref{sec:N4}. In Section \ref{sec:A6}, we turn to the $\mathcal{A}(6)$ VOA, demonstrate the failure of classical freeness, and determine the generalized arc space which reproduces the Schur and Macdonald indices of $\widehat{E}_6(SU(2))$. We conclude in Section \ref{sec:disc} with a discussion of open questions.

\section{Generalized arc spaces}\label{sec:arcs}

In this section, we review the construction of the jet schemes and the arc space associated to a ring, and we introduce the generalization that is necessary for this paper. Since the rings that we consider have both bosonic and fermionic generators, we work with $\mathbb{Z}_2$-graded-commutative rings from the start; the commutative case is recovered when there are no fermionic generators.

Let $S$ be the $\mathbb{Z}_2$-graded-commutative polynomial ring with bosonic generators $x_1, \cdots, x_m$ and fermionic generators $y_1, \cdots, y_r$:
\begin{equation}
    S = \mathbb{C}[x_1, \cdots, x_m] \otimes \Lambda[y_1, \cdots, y_r] \,.
\end{equation}
Any finitely-generated $\mathbb{Z}_2$-graded-commutative ring can be written as a quotient $R = S \, / \, I$, for some ideal $I$ of $S$. We equip $S$ with a multigrading, specified by the degree assignments
\begin{equation}
    \operatorname{deg}(x_i) = (\alpha_i, \beta_i, \cdots) \,, \qquad \operatorname{deg}(y_a) = (\gamma_a, \kappa_a, \cdots) \,.
\end{equation}
In our applications, the first grading is the $q$-grading, given by the conformal dimension, the second is the $T$-grading, and any further gradings correspond to flavor symmetries. Every generator has positive $q$-degree.

Two further pieces of notation will be useful. For $f \in S$, we write $\operatorname{Jac}(f)$ for the Jacobian ideal of $f$, that is, the ideal of $S$ generated by the partial derivatives of $f$ with respect to all of the coordinates, both bosonic and fermionic.\footnote{For a fermionic coordinate, the left and right derivatives of a polynomial of definite fermion number differ by an overall sign, and so the ideal does not depend on this choice.} We also use odd derivations of $S$, that is, $\mathbb{C}$-linear maps, $\delta$, which exchange the bosonic and fermionic elements of $S$ and satisfy $\delta(fg) = (\delta f) g + (-1)^{|f|} f (\delta g)$, where $|f|$ is the fermion number of $f$; such a derivation is determined by its action on the coordinates.

To describe the arc space, we replace each generator by a formal power series in an auxiliary variable, $t$:
\begin{equation}
    x_i(t) = \sum_{j \geq 0} x_i^{(j)} t^j \,, \qquad y_a(t) = \sum_{j \geq 0} y_a^{(j)} t^j \,,
\end{equation}
where $x_i^{(j)}$ are bosonic and $y_a^{(j)}$ are fermionic. Let $S_\infty$ denote the $\mathbb{Z}_2$-graded-commutative polynomial ring in all of the $x_i^{(j)}$ and $y_a^{(j)}$, with $j \geq 0$, and let $S_n \subset S_\infty$ denote the subring generated by those with $j \leq n$; thus, $S_n$ has $m(n+1)$ bosonic and $r(n+1)$ fermionic generators. We identify $S$ with $S_0$ via $x_i = x_i^{(0)}$ and $y_a = y_a^{(0)}$. The multigrading on $S$ induces a multigrading on $S_\infty$, with one additional grading, the jet-grading:
\begin{equation}
    \operatorname{deg}(x_i^{(j)}) = (j, \alpha_i, \beta_i, \cdots) \,, \qquad \operatorname{deg}(y_a^{(j)}) = (j, \gamma_a, \kappa_a, \cdots) \,.
\end{equation}
Thus the $q$-degree of $x_i^{(j)}$ is that of $x_i$, while the jet-degree counts the derivatives; the conformal dimension of $x_i^{(j)}$ is the sum of the two, $\alpha_i + j$. The ring $S_\infty$ admits an even derivation, $D$, which acts as $d/dt$ on the formal power series:
\begin{equation}\label{eqn:derivationDef}
    D x_i^{(j)} = (j+1) x_i^{(j+1)} \,, \qquad D y_a^{(j)} = (j+1) y_a^{(j+1)} \,,
\end{equation}
and which is extended to all of $S_\infty$ by $\mathbb{C}$-linearity and the Leibniz rule; since $D$ is even, the Leibniz rule involves no signs. The derivation raises the jet-degree by one, and preserves all of the other degrees. With this normalization, $x_i^{(j)}$ corresponds to $\partial^j x_i / j!$, and, for any $f \in S$, the coefficient of $t^\ell$ in $f(x(t), y(t))$ is $D^\ell f / \ell!$. In Sections \ref{sec:vec} and \ref{sec:A6}, where we label the generators by individual letters, we write, for example, $x_j$ for $x^{(j)}$; elsewhere, subscripts on the coordinates label their components, such as flavor weights, rather than derivatives.

Let $I = (f_1, \cdots, f_s)$ be an ideal of $S$, and $R = S \, / \, I$. The $n$th jet ring of $R$ is obtained by imposing the coefficients of $t^0, \cdots, t^n$ in each relation $f_c(x(t), y(t))$:
\begin{equation}
    J_n(R) = S_n \, / \, D^{\leq n}I \,, \qquad D^{\leq n}I = \left( D^\ell f_c \,\, \big| \,\, c = 1, \cdots, s \,;\, 0 \leq \ell \leq n \right) \,.
\end{equation}
Since $D^\ell f_c$ only involves the $x_i^{(j)}$ and $y_a^{(j)}$ with $j \leq \ell$, these generators indeed belong to $S_n$; furthermore, the Leibniz rule implies that $D^{\leq n}I$ does not depend on the choice of generators of $I$. The coordinate ring of the arc space is the direct limit
\begin{equation}
    J_\infty(R) = \varinjlim_n \, J_n(R) = S_\infty \, / \, DI \,, \qquad DI = \left( D^\ell f_c \,\, \big| \,\, c = 1, \cdots, s \,;\, \ell \geq 0 \right) \,,
\end{equation}
where $DI$ is the differential ideal generated by $I$.

In the language of differential algebra \cite{Ritt,Kolchin}, $S_\infty$, together with $D$, is the ring of differential polynomials in the $x_i$ and $y_a$, and $J_\infty(R)$ is its quotient by the differential ideal generated by relations of order zero, that is, relations which do not involve derivatives. The generalization that we require allows relations which involve derivatives up to some finite order, $k$. Let $I_k$ be an ideal of $S_k$. Then, for $n \geq k$, we define
\begin{equation}\label{eqn:knJet}
    J_{n,k}(S, I_k) = S_n \, / \, D^{\leq n-k}I_k \,, \qquad J_{\infty,k}(S, I_k) = \varinjlim_n \, J_{n,k}(S, I_k) = S_\infty \, / \, DI_k \,.
\end{equation}
As before, $D^\ell g$ involves only the $x_i^{(j)}$ and $y_a^{(j)}$ with $j \leq k + \ell$, for any $g \in S_k$. When $k = 0$, this reduces to the jet rings and arc space described above: $J_{n,0}(S, I) = J_n(S \, / \, I)$ and $J_{\infty,0}(S, I) = J_\infty(S \, / \, I)$. More generally, if $I_k = D^{\leq k}I$ for an ideal $I$ of $S$, then $J_{\infty,k}(S, I_k) = J_\infty(S \, / \, I)$.

This generalization is motivated by the VOA. Let $R_\mathbb{V} = S \, / \, I$, where the generators of $S$ correspond to the strong generators of $\mathbb{V}$. The composition of the natural projection $S_\infty \rightarrow J_\infty(R_\mathbb{V})$ with the map $\Phi$ in equation \eqref{eqn:CF} is a surjective homomorphism of differential algebras, $S_\infty \rightarrow \operatorname{gr}_{\operatorname{Li}}(\mathbb{V})$, and its kernel, $K$, is a differential ideal which contains $DI$. The VOA is classically free if and only if $K = DI$, that is, if and only if $K$ is generated, as a differential ideal, by relations of order zero. Equation \eqref{eqn:biggereqn} instead allows $K$ to be generated by relations of order at most one.

A priori, it is not clear that $K$ is generated by relations of any finite order. In fact, $K$ is generated by relations of order at most $k$, for some $k$, if and only if it is finitely generated as a differential ideal: a finite set of generators lies in $S_k$ for some $k$, and, conversely, if $K$ is generated by relations of order at most $k$, then it is generated by the ideal $K \cap S_k$ of $S_k$, which is finitely generated by Hilbert's basis theorem. However, there is no analogue of Hilbert's basis theorem for differential ideals: even in the ring of differential polynomials in a single variable, there are differential ideals which are not finitely generated \cite{Ritt}. The basis theorem of Ritt and Raudenbush only guarantees that every radical differential ideal is the radical of a finitely-generated differential ideal \cite{Ritt,Kolchin}, and $K$ is, in general, not radical; for example, the images of the fermionic generators in $\operatorname{gr}_{\operatorname{Li}}(\mathbb{V})$ are non-zero and nilpotent. Therefore, the existence of a generalized arc space that is isomorphic to $\operatorname{gr}_{\operatorname{Li}}(\mathbb{V})$ is a non-trivial property of the VOA; indeed, whether $K$ is finitely generated was raised as a question in \cite{ArakawaLinshaw:2019}, for a VOA which is not classically free. Nevertheless, $K$ is finitely generated in every example that we are aware of. For the Ising model, which is not classically free, $K$ is generated by $DI$ together with a single relation of order three \cite{MR4589087}, and, for the VOAs studied in this paper, relations of order at most one reproduce the vacuum character to the order that we have checked. We therefore conjecture that, for every VOA arising from a 4d $\mathcal{N}=2$ SCFT, $K$ is finitely generated as a differential ideal, that is, $\operatorname{gr}_{\operatorname{Li}}(\mathbb{V}) \cong J_{\infty,k}(S, I_k)$ for some finite $k$ and some ideal $I_k$ of $S_k$.

Finally, we describe the structure that enters the Hilbert series. We take $I_k$ to be generated by polynomials which are homogeneous with respect to the jet-grading and the $q$-grading, but not necessarily with respect to the $T$-grading. Then $J_{\infty,k}(S, I_k)$ inherits the jet-grading and the $q$-grading, while the $T$-grading on $S_\infty$ induces only an increasing, multiplicative $T$-filtration: $F_s J_{\infty,k}(S, I_k)$ is spanned by the images of the monomials of $T$-degree at most $s$. Since $D$ preserves the $T$-degree, the filtration is compatible with the derivation. The associated graded ring,
\begin{equation}
    \operatorname{gr}(J_{\infty,k}(S, I_k)) = \bigoplus_s F_s J_{\infty,k}(S, I_k) \, / \, F_{<s} J_{\infty,k}(S, I_k) \,,
\end{equation}
is the quotient of $S_\infty$ by the ideal spanned by the components of highest $T$-degree of the elements of $DI_k$, and it can be computed from a Gr\"obner basis with respect to a monomial order which refines the $T$-degree. We emphasize that the associated graded ring is taken after forming the arc space; in general, $\operatorname{gr}(J_\infty(R))$ is not isomorphic to $J_\infty(\operatorname{gr}(R))$. The Hilbert series of a ring, $A$, with this structure is
\begin{equation}
    \operatorname{HS}_{p,q,T,\tau}(A) = \sum_{a,b,c,\epsilon} \dim(A_{a,b,c,\epsilon}) \, p^a q^b T^c \tau^\epsilon \,,
\end{equation}
where $A_{a,b,c,\epsilon}$ is the subspace of jet-degree $a$, $q$-degree $b$, $T$-degree $c$, and fermion number $\epsilon \in \{0, 1\}$; each such subspace is finite-dimensional, since every generator has positive $q$-degree. Additional fugacities can be included for flavor gradings. We write $\operatorname{HS}_{q,q,T,-1}$ for the specialization to $p = q$ and $\tau = -1$, and $\operatorname{HS}_{q,q,-1}$ when, in addition, the $T$-filtration is forgotten.

\section{The free \texorpdfstring{\boldmath{$\mathcal{N}=2$}}{N=2} vector}\label{sec:vec}

The simplest 4d $\mathcal{N}=2$ SCFT whose Schur sector contains fermionic operators is the free $\mathcal{N}=2$ vector multiplet. Its Macdonald index is \cite{Gadde:2011uv}
\begin{equation}\label{eqn:freebigmac}
    I_\text{Mac}^{\mathcal{N}=2\text{ vector}}(q, T) = \operatorname{PE} \left[ \frac{-q-qT}{1-q} \right] \,.
\end{equation}
As discussed in Section \ref{Intro}, the expansion of this index contains negative coefficients, and so there is no commutative filtered graded ring, $R$, which reproduces it via equation \eqref{eqn:bigeqn}. However, this example requires only the most direct extension of the proposal: the ring is a free $\mathbb{Z}_2$-graded-commutative ring, and the Hilbert series of its arc space reproduces the Macdonald index exactly.

The $\mathcal{N}=2$ vector multiplet contains the gauge field, $A_\mu$, the gaugini, $\lambda^I_\alpha$ and $\widetilde{\lambda}_{I\dot{\alpha}}$, which are doublets of $\mathfrak{su}(2)_R$, and a complex scalar, $\phi$. The Schur operators are generated by the gaugino components $\lambda^1_+$ and $\widetilde{\lambda}_{1\dot{+}}$, together with their derivatives, and the 4d free field OPEs determine the associated VOA to be the small algebra of the $bc$-system of weights $(1, 0)$, that is, the subalgebra generated by $b$ and $\partial c$ \cite{Beem:2013sza}. Under the SCFT/VOA correspondence, $\lambda^1_+ \mapsto b$ and $\widetilde{\lambda}_{1\dot{+}} \mapsto \partial c$. The 4d charges of these operators, and the resulting $q$- and $T$-degrees, are listed in Table \ref{tbl:vec}.

\begin{table}[H]
    \centering
    \begin{threeparttable}
    \begin{tabular}{cccccccc}
        \toprule
        4d operator & 2d field & $\Delta$ & $(j_1, j_2)$ & $R$ & $r$ & $\Delta - R$ & $R - r$ \\\midrule
        $\lambda^1_+$ & $b$ & $\frac{3}{2}$ & $(\frac{1}{2}, 0)$ & $\frac{1}{2}$ & $-\frac{1}{2}$ & $1$ & $1$ \\[3pt]
        $\widetilde{\lambda}_{1\dot{+}}$ & $\partial c$ & $\frac{3}{2}$ & $(0, \frac{1}{2})$ & $\frac{1}{2}$ & $\frac{1}{2}$ & $1$ & $0$ \\\bottomrule
    \end{tabular}
    \end{threeparttable}
    \caption{The Schur operators which generate the Schur sector of the free $\mathcal{N}=2$ vector multiplet, and the corresponding fields of the small $bc$-system. We list the conformal dimension ($\Delta$), the Lorentz spins ($j_1, j_2$), and the $\mathfrak{su}(2)_R$ and $\mathfrak{u}(1)_r$ charges ($R$, $r$); the last two columns are the $q$-degree and the $T$-degree, respectively.}
    \label{tbl:vec}
\end{table}

We therefore take the ring to have one fermionic generator for each of $b$ and $\partial c$, with the $q$- and $T$-degrees given by the last two columns of Table \ref{tbl:vec}:
\begin{equation}
    R = S = \Lambda[y, z] \,, \qquad \operatorname{deg}(y) = (1, 1) \,, \quad \operatorname{deg}(z) = (1, 0) \,,
\end{equation}
where the first entry is the $q$-degree and the second is the $T$-degree. There are no relations, that is, the ideal $I_0$ of $S$ vanishes. In particular, the $T$-grading is an honest grading, and passing to the associated graded ring has no effect. We use $\partial c$, rather than $c$, as a generator since $c$ has conformal dimension zero; including it would violate the requirement of Section \ref{sec:arcs} that every generator has positive $q$-degree, and the graded components of the arc space would no longer be finite-dimensional. Since the fermionic generators are nilpotent, $\operatorname{Spec}(R)_\text{red}$ is a point, consistent with the trivial Higgs branch of the free vector multiplet. Furthermore, $R = R_\mathbb{V}$, and the VOA is classically free: the kernel, $K$, of the map $S_\infty \rightarrow \operatorname{gr}_{\operatorname{Li}}(\mathbb{V})$ described in Section \ref{sec:arcs} vanishes.

The coordinate ring of the arc space is then the free $\mathbb{Z}_2$-graded-commutative ring $J_\infty(R) = S_\infty = \Lambda[y_0, y_1, \cdots, z_0, z_1, \cdots]$, where $y_j$ and $z_j$ correspond to $\partial^j b / j!$ and $\partial^{j+1} c / j!$, respectively. Its Hilbert series is
\begin{equation}
    \operatorname{HS}_{p,q,T,\tau}(J_\infty(R)) = \prod_{j = 0}^\infty (1 + \tau p^j q T)(1 + \tau p^j q) = (-qT\tau; p)_\infty (-q\tau; p)_\infty \,,
\end{equation}
where the first factor in the product comes from the $y_j$, and the second from the $z_j$, and where we have written the product in terms of the $q$-Pochhammer symbol, $(a; p)_n = \prod_{i=0}^{n-1} (1 - a p^i)$, with $n = \infty$. Setting $p = q$ and $\tau = -1$, we find
\begin{equation}
    \operatorname{HS}_{q,q,T,-1}(J_\infty(R)) = (qT; q)_\infty (q; q)_\infty = I_\text{Mac}^{\mathcal{N}=2\text{ vector}}(q, T) \,,
\end{equation}
in agreement with equation \eqref{eqn:freebigmac}. In the Schur limit, $T \rightarrow 1$, this becomes $(q; q)_\infty^2$, which is, up to the factor $q^{-c/24}$, the vacuum character of the small $bc$-system.

\section{\texorpdfstring{\boldmath{$\mathcal{N}=4$}}{N=4} SYM}\label{sec:N4}

After considering the free vector multiplet, we turn to, perhaps, the most canonical example of an interacting 4d $\mathcal{N}=2$ SCFT where the chiral algebra has fermionic strong generators: $\mathcal{N}=4$ super Yang--Mills.\footnote{See \cite{Bonetti:2025kan,Gaberdiel:2025eaf} for detailed studies of the VOA of $\mathcal{N}=4$ super Yang--Mills.} For gauge algebra $\mathfrak{g} = \mathfrak{su}(N)$, the generators of the chiral algebra are expected to be the components of $N-1$ small $\mathcal{N}=4$ multiplets, one for each degree $p = 2, \cdots, N$ of the Casimirs, the $p$th of which has $4p$ components; the components of the $p = 2$ multiplet generate the small $\mathcal{N}=4$ superconformal algebra with $c = -3(N^2 - 1)$ \cite{Beem:2013sza}. In this section, we study the cases $\mathfrak{g} = \mathfrak{su}(2)$ and $\mathfrak{g} = \mathfrak{su}(3)$ in turn.

In both cases, we can determine the Macdonald index using the fact that we are considering a Lagrangian gauge theory. The Macdonald index of the free vector multiplet with gauge algebra $\mathfrak{g}$ is a simple generalization of equation \eqref{eqn:freebigmac}, to wit,
\begin{equation}
    I_\text{Mac}^{\text{vec}}(q, T, z) = \operatorname{PE} \left[ \frac{-q - qT}{1-q} \chi_{\text{adj}}(z) \right] \,,
\end{equation}
where $\chi_{\text{adj}}(z)$ is the character of the adjoint representation of $\mathfrak{g}$, and $z$ denotes the fugacities for its Cartan subalgebra; for $\mathfrak{g} = \mathfrak{su}(2)$, $\chi_\text{adj}(z) = \chi_{\bm{3}}(z)$, where $\chi_{\bm{d}}(z)$ is the character of the $d$-dimensional representation of $\mathfrak{su}(2)$. Similarly, it is known that the Macdonald index of a $\mathfrak{g}$-adjoint-valued $\mathcal{N}=2$ (full) hypermultiplet is
\begin{equation}\label{eqn:N4Mac}
    I_\text{Mac}^{\text{hyper}}(q, T, z, a) = \operatorname{PE} \left[ \frac{(q T)^{\frac{1}{2}}}{1-q} \chi_{\bm{2}}(a) \chi_{\text{adj}}(z) \right] \,,
\end{equation}
where $a$ is the fugacity for the $\mathfrak{su}(2)$ flavor symmetry. Therefore, the Macdonald index of the $\mathcal{N}=4$ theory can be straightforwardly worked out:
\begin{equation}\label{eqn:N4Macdonald}
    I_\text{Mac}^{\mathcal{N}=4}(q, T, a) = \int [dz] I_\text{Mac}^{\text{vec}}(q, T, z) I_\text{Mac}^{\text{hyper}}(q, T, z, a) \,,
\end{equation}
where $[dz]$ is the Haar measure for $\mathfrak{g}$.

\subsection{Gauge algebra \texorpdfstring{$\mathfrak{su}(2)$}{su(2)}}\label{sec:N4su2}

For $\mathfrak{g} = \mathfrak{su}(2)$, the chiral algebra is believed to be the small $\mathcal{N}=4$ superconformal algebra with $c = -9$ \cite{Beem:2013sza}. There are seven strong generators, which we list in Table \ref{tbl:N4SG}.

The known, low-lying null relations of this VOA (or, equivalently, the OPE decoupling relations in 4d) can be written, keeping only the terms that survive in $R_{\mathbb{V}}$, as the vanishing of the following expressions \cite{MR3492038,Bonetti:2025kan}:
\begin{equation}\label{eqn:N4nulls}
\begin{gathered}
    J_+ G_- + J_0 G_+ \,, \quad J_- G_+ - J_0 G_- \,, \quad 
    -J_+ \widetilde{G}_- + J_0 \widetilde{G}_+ \,, \quad - J_- \widetilde{G}_+ - J_0 \widetilde{G}_- \,, \quad G_+ G_- \,, \\ \widetilde{G}_+ \widetilde{G}_- \,, \quad 
    G_+ \widetilde{G}_- + J_0 T \,, \quad G_- \widetilde{G}_+ - J_0 T \,, \quad 
    G_+ \widetilde{G}_+ + J_+ T \,, \quad G_- \widetilde{G}_- + J_- T \,.
\end{gathered}
\end{equation}
We have used that the stress tensor, $T$, is the Sugawara stress tensor, that is,
\begin{equation}\label{eqn:N4sugawara}
    T = J_+ J_- + J_0^2 \,,
\end{equation}
where the $J_{\pm}$, $J_0$ are the currents in the adjoint of the additional $\mathfrak{su}(2)$ global symmetry.

Based on this collection of strong generators and null relations, we are led to consider the $\mathbb{Z}_2$-graded-commutative ring
\begin{equation}
    S = \mathbb{C}[j, j_+, j_-, t] \otimes \Lambda[g_+, g_-, \widetilde{g}_+, \widetilde{g}_-] \,,
\end{equation}
where the coordinates correspond to the strong generators of the VOA, together with the stress tensor, as
\begin{equation}
    (j, j_+, j_-, t, g_+, g_-, \widetilde{g}_+, \widetilde{g}_-) \leftrightarrow (J_0, J_+, J_-, T, G_+, G_-, \widetilde{G}_+, \widetilde{G}_-) \,.
\end{equation}
The degrees of all of the coordinates other than $t$ are as in Table \ref{tbl:N4SG}. The coordinate $t$ is special: the stress tensor is not a strong generator, since it is given by the Sugawara construction as in equation \eqref{eqn:N4sugawara}; however, its $T$-degree is not the one implied by this relation, and instead we must assign
\begin{equation}
    \operatorname{deg}(t) = (2, 1, 0) \,,
\end{equation}
where we list the $q$-degree, the $T$-degree, and the $\mathfrak{su}(2)$ flavor charge, in that order. We consider the following ideal of $S$, which captures the null relations in equation \eqref{eqn:N4nulls}, as well as the Sugawara relation, which is not homogeneous in the $T$-grading:
\begin{equation}\label{eqn:N4ideal}
  \begin{aligned}
    I \,\,=\,\, \big( & j_+ g_- + j g_+ \,, \,\, j_- g_+ - j g_- \,, \,\, -j_+ \widetilde{g}_- + j \widetilde{g}_+ \,, \,\, -j_- \widetilde{g}_+ - j \widetilde{g}_- \,, \,\, g_+ g_- \,, \,\, \widetilde{g}_+ \widetilde{g}_- \,, \\ & g_+ \widetilde{g}_- + j t \,, \,\, g_- \widetilde{g}_+ - j t \,, \,\, g_+ \widetilde{g}_+ + j_+ t \,, \,\, g_- \widetilde{g}_- + j_- t \,, \,\, t - j_+ j_- - j^2 \big) \,.
  \end{aligned}
\end{equation}
We will see in Section \ref{sec:N3} that this ideal, including the Sugawara relation, has a more compact origin: it is the $p = 2$ member of a family of ideals, built from the derivatives of a superpotential, which describe all the rank-one $\mathcal{N}\geq 3$ SCFTs.

\begin{table}[H]
    \centering
    \begin{threeparttable}
    \begin{tabular}{cccccccc}
        \toprule
        Strong Gen. & $J_0$ & $J_+$ & $J_-$ & $G_+$ & $G_-$ & $\widetilde{G}_+$ & $\widetilde{G}_-$ \\\midrule
        $F$ & $0$ & $0$ & $0$ & $1$ & $1$ & $1$ & $1$ \\[1pt]
        $h$ & $1$ & $1$ & $1$ & $\frac{3}{2}$ & $\frac{3}{2}$ & $\frac{3}{2}$ & $\frac{3}{2}$ \\[3pt]
        $a$ & $0$ & $2$ & $-2$ & $1$ & $-1$ & $1$ & $-1$ \\[3pt]
        $T$ & $1$ & $1$ & $1$ & $\frac{1}{2}$ & $\frac{1}{2}$ & $\frac{3}{2}$ & $\frac{3}{2}$ \\\bottomrule
    \end{tabular}
    \end{threeparttable}
    \caption{The strong generators of the chiral algebra associated to $\mathcal{N}=4$ super Yang--Mills with gauge algebra $\mathfrak{su}(2)$. We list the fermion number ($F$), the conformal dimension ($h$), the charge under the $\mathfrak{su}(2)$ flavor symmetry $(a)$, and, finally, the $T$-grading $(T)$.}
    \label{tbl:N4SG}
\end{table}

We can then compute the Hilbert series of the arc space $J_\infty(S \, / \, I) = J_{\infty, 0}(S, I)$ and compare to the expressions for the Schur and Macdonald indices of the SYM theory. First, we consider the Schur limit, where we forget about the $T$-grading, and we find:\footnote{We have computed these Hilbert series to much higher orders, as we discuss below, but it is unilluminating to write them all here explicitly. The full expressions, including the fugacity of the $\mathfrak{su}(2)$ global symmetry, are given in the {\tt Mathematica} notebook {\tt ArcSpaceHS.nb} attached to the arXiv submission of this paper.}
\begin{equation}\label{eqn:N4HSS}
  \begin{aligned}
    \operatorname{HS}_{q,q,-1}&(J_{\infty}(S \, / \, I)) = 1 + 3 q - 4 q^\frac{3}{2} + 9 q^2 - 12 q^\frac{5}{2} + 22 q^3 - 36 q^\frac{7}{2} + 
 60 q^4 - 88 q^\frac{9}{2} \\&+ 135 q^5 - 204 q^\frac{11}{2} + 302 q^6 -432 q^\frac{13}{2} + 627 q^7 - 900 q^\frac{15}{2} + 1269 q^8 \\&-1764 q^\frac{17}{2} + 2451 q^9 - 3384 q^\frac{19}{2} + 4629 q^{10} - 6268q^\frac{21}{2} + 8460q^{11} + \cdots \,.
  \end{aligned}
\end{equation}
Similarly, we can determine the Hilbert series with the $T$-grading, and we find
\begin{equation}\label{eqn:N4HS}
\begin{aligned}
    &\operatorname{HS}_{q,q,T,-1}(\operatorname{gr}(J_{\infty}(S \, / \,  I))) = 1 + 3qT
- 2q^{\frac{3}{2}}\left(T^{\frac{1}{2}} + T^{\frac{3}{2}}\right)
\\&\qquad+ q^2\left(4T + 5T^2\right)
 - 2q^{\frac{5}{2}}\left(T^{\frac{1}{2}} + 3T^{\frac{3}{2}} + 2T^{\frac{5}{2}}\right)
+ q^3\left(4T + 11T^2 + 7T^3\right)
\\&\qquad- 2q^{\frac{7}{2}}\left(T^{\frac{1}{2}} + 6T^{\frac{3}{2}} + 8T^{\frac{5}{2}} + 3T^{\frac{7}{2}}\right)
+ q^4\left(7T + 24T^2 + 20T^3 + 9T^4\right)
\\&\qquad- 2q^{\frac{9}{2}}\left(T^{\frac{1}{2}} + 10T^{\frac{3}{2}} + 17T^{\frac{5}{2}} + 12T^{\frac{7}{2}} + 4T^{\frac{9}{2}}\right)
\\&\qquad+ q^5\left(8T + 39T^2 + 49T^3 + 28T^4 + 11T^5\right)
\\&\qquad - 2q^{\frac{11}{2}}\left(T^{\frac{1}{2}} + 14T^{\frac{3}{2}} + 34T^{\frac{5}{2}} + 32T^{\frac{7}{2}} + 16T^{\frac{9}{2}} + 5T^{\frac{11}{2}}\right)
\\&\qquad + q^6\left(11T + 61T^2 + 105T^3 + 76T^4 + 36T^5 + 13T^6\right) \\&\qquad - 2q^\frac{13}{2}\left(T^\frac{1}{2} + 18T^\frac{3}{2} + 57T^\frac{5}{2} + 70T^\frac{7}{2} + 44T^\frac{9}{2} + 20T^\frac{11}{2} + 6T^\frac{13}{2}\right)
\\&\qquad + q^7\left(12T + 88T^2 + 189T^3 + 179T^4 + 100T^5 + 44T^6 + 15T^7 \right)
\\&\qquad - 2q^\frac{15}{2}\left(T^\frac{1}{2} + 24T^\frac{3}{2} + 92T^\frac{5}{2} + 138T^\frac{7}{2} + 108T^\frac{9}{2} + 56T^\frac{11}{2} + 24T^\frac{13}{2} + 7T^\frac{15}{2}\right)
\\&\qquad + q^8\left(15T + 125T^2 + 320T^3 + 368T^4 + 248T^5 + 124T^6 + 52T^7 + 17T^8 \right)
+ \cdots \,.
  \end{aligned}
\end{equation}
We can see that equations \eqref{eqn:N4HSS} and \eqref{eqn:N4HS} match with the series expansion of the Macdonald index as in equation \eqref{eqn:N4Macdonald}. In fact, we have computed the Hilbert series with both the $T$-grading and the fugacity, $a$, of the $\mathfrak{su}(2)$ global symmetry turned on, up to and including order $q^{21}$, and it agrees with the series expansion of equation \eqref{eqn:N4Macdonald} at every order, that is, for the first forty-three coefficients of the expansion in powers of $q^{1/2}$, each of which is a polynomial in $T^{1/2}$ and $a^{\pm 1}$. Since the ideal in equation \eqref{eqn:N4ideal} is an ideal of $S$ (i.e., it was not necessary to consider an ideal of $S_1$ that is not an ideal of $S$), and since $S \, / \, I$ is isomorphic to $R_\mathbb{V}$ once $t$ is eliminated using the Sugawara relation, this matching of the Hilbert series and the (unrefined) character of the VOA indicates that the small $\mathcal{N}=4$ superconformal algebra with $c = -9$ is classically free.

\begin{table}[H]
    \centering
    \begin{threeparttable}
    \begin{tabular}{cccccc}
        \toprule
        Generator & Coordinate & $F$ & $h$ & $a$ & $T$ \\\midrule
        $T$ & $t$ & $0$ & $2$ & $0$ & $1$ \\[1pt]
        $J_0$, $J_\pm$ & $j$, $j_\pm$ & $0$ & $1$ & $0$, $\pm 2$ & $1$ \\[1pt]
        $G_\pm$ & $g_\pm$ & $1$ & $\frac{3}{2}$ & $\pm 1$ & $\frac{1}{2}$ \\[3pt]
        $\widetilde{G}_\pm$ & $\widetilde{g}_\pm$ & $1$ & $\frac{3}{2}$ & $\pm 1$ & $\frac{3}{2}$ \\[3pt]
        $W_m$ & $w_m$ & $0$ & $\frac{3}{2}$ & $\pm 1$, $\pm 3$ & $\frac{3}{2}$ \\[3pt]
        $W^+_m$ & $\psi_m$ & $1$ & $2$ & $0$, $\pm 2$ & $1$ \\[1pt]
        $W^-_m$ & $\chi_m$ & $1$ & $2$ & $0$, $\pm 2$ & $2$ \\[1pt]
        $U_m$ & $u_m$ & $0$ & $\frac{5}{2}$ & $\pm 1$ & $\frac{3}{2}$ \\\bottomrule
    \end{tabular}
    \end{threeparttable}
    \caption{The generators of the chiral algebra associated to $\mathcal{N}=4$ super Yang--Mills with gauge algebra $\mathfrak{su}(3)$, and the corresponding coordinates of $S_0$. The first four rows form the $p = 2$ multiplet, and the last four the $p = 3$ multiplet. We list the fermion number ($F$), the conformal dimension ($h$), the charge under the $\mathfrak{su}(2)$ flavor symmetry ($a$), and the $T$-grading ($T$); the subscript $m$ labels the components of each $\mathfrak{su}(2)$ multiplet by their weight, so that $a = 2m$. The doublet $U_m$ is not a strong generator; see equation \eqref{eqn:N4SU3r}.}
    \label{tbl:N4SU3SG}
\end{table}

\subsection{Gauge algebra \texorpdfstring{$\mathfrak{su}(3)$}{su(3)}}\label{sec:N4su3}

For $\mathfrak{g} = \mathfrak{su}(3)$, there are two multiplets, with $p = 2$ and $p = 3$, and thus twenty generators, which we list in Table \ref{tbl:N4SU3SG}. The components of the $p = 2$ multiplet generate the small $\mathcal{N}=4$ superconformal algebra with $c = -24$, and we denote them as in the $\mathfrak{su}(2)$ case; unlike for $\mathfrak{su}(2)$, the stress tensor, $T$, is not the Sugawara stress tensor. The $p = 3$ multiplet consists of a bosonic $\mathfrak{su}(2)$ quadruplet, $W_m$, of conformal dimension $3/2$, which is its bottom component, two fermionic $\mathfrak{su}(2)$ triplets, $W^+_m$ and $W^-_m$, of conformal dimension two, and a bosonic $\mathfrak{su}(2)$ doublet, $U_m$, of conformal dimension $5/2$, which is its top component. Here, and in the following, the subscript $m$ is the weight under the $\mathfrak{su}(2)$ flavor symmetry, so that the flavor charge is $a = 2m$. The $T$-degrees of the stress tensor and of the $p = 2$ multiplet are as for $\mathfrak{su}(2)$. For the $p = 3$ multiplet, the bottom and top components both have $T$-degree $3/2$, while $W^+_m$, which is the triplet that appears in the OPE of $W_m$ with the supercurrents $G_\pm$ of $T$-degree $1/2$, has $T$-degree one, and $W^-_m$ has $T$-degree two. We are thus led to consider the $\mathbb{Z}_2$-graded-commutative ring
\begin{equation}
    S_0 = \mathbb{C}[t, j, j_\pm, w_m, u_m] \otimes \Lambda[g_\pm, \widetilde{g}_\pm, \psi_m, \chi_m] \,,
\end{equation}
where the coordinates correspond to the generators of the VOA as in Table \ref{tbl:N4SU3SG}, and have the degrees listed there; their normalizations, which are fixed by the OPEs that follow from the free-field realization of the VOA \cite{Bonetti:2018fqz}, are those of the {\tt Mathematica} file {\tt N4SU3.m} attached to the arXiv submission.

We have computed the null relations of the VOA from its free-field realization \cite{Bonetti:2018fqz}, up to and including conformal dimension five. Keeping only the terms that survive in $R_\mathbb{V}$, we obtain an ideal $I_0$ of $S_0$, and we find that, up to this conformal dimension, $I_0$ has fifty-nine minimal generators: two at conformal dimension $5/2$, eleven at conformal dimension three, twenty-four at conformal dimension $7/2$, twenty-two at conformal dimension four, and none at conformal dimensions $9/2$ and five. The two minimal generators at conformal dimension $5/2$, which have flavor charge $\pm 1$, are
\begin{equation}\label{eqn:N4SU3r}
  \begin{aligned}
    r_- &= u_{-1/2} - \frac{1}{3} j w_{-1/2} + \frac{1}{3} j_+ w_{-3/2} + \frac{2}{3} j_- w_{1/2} \,, \\
    r_+ &= u_{1/2} - \frac{2}{3} j w_{1/2} + \frac{1}{3} j_+ w_{-1/2} + 2 j_- w_{3/2} \,.
  \end{aligned}
\end{equation}
They are linear in $u_{\pm 1/2}$: as for the Sugawara stress tensor of the $\mathfrak{su}(2)$ theory, they express the top component of the $p = 3$ multiplet in terms of the other generators, and thus the VOA has eighteen strong generators. We have listed the fifty-nine minimal generators of $I_0$, together with the degrees of the coordinates, in the file {\tt N4SU3.m}.

Let $R = S_0 \, / \, I_0$. We have computed the Hilbert series of $\operatorname{gr}(J_\infty(R))$, with both the $T$-grading and the fugacity of the $\mathfrak{su}(2)$ global symmetry turned on, up to and including order $q^{12}$, and it agrees with the series expansion of the Macdonald index in equation \eqref{eqn:N4Macdonald}, with $\mathfrak{g} = \mathfrak{su}(3)$, at every order, that is, for the first twenty-five coefficients of the expansion in powers of $q^{1/2}$; in particular, the Schur limit agrees with the Schur index up to and including order $q^{12}$. Both series are given in the {\tt Mathematica} notebook {\tt ArcSpaceHS.nb}. Since this agreement extends well beyond conformal dimension four, it indicates that the null relations up to conformal dimension four generate the ideal, at least up to this order, and, since $I_0$ is an ideal of $S_0$, that the VOA is classically free. We have also verified this directly up to and including conformal dimension five: for each null relation, the component with the lowest number of derivatives is an element of the kernel, $K$, of the map $S_\infty \rightarrow \operatorname{gr}_{\operatorname{Li}}(\mathbb{V})$ described in Section \ref{sec:arcs}, and we find that these components span exactly $DI_0$ at each conformal dimension, flavor charge, and fermion number.

The ideal $I_0$ also has a compact presentation in terms of superpotentials, via the construction that we describe in Section \ref{sec:N3} for the rank-one $\mathcal{N}=3$ SCFTs. Let $\delta_+$ and $\delta_-$ be the supersymmetry variations associated with the supercurrents $\widetilde{G}_+$ and $G_-$ of the $p = 2$ multiplet, in the notation of Table \ref{tbl:N4SU3SG}, which are defined, as in equation \eqref{eqn:N3superfields}, by the expansions of the $\mathcal{N}=2$ superfields that contain the generators. They act on the coordinates as
\begin{equation}\label{eqn:N4SU3delta}
  \begin{aligned}
    \delta_+ j &= -\widetilde{g}_+ \,, \quad \delta_+ j_- = -\widetilde{g}_- \,, \quad \delta_+ g_- = -t \,, \\
    \delta_+ w_m &= -\left(\tfrac{3}{2} - m\right) \chi_{m + \frac{1}{2}} \,, \quad \delta_+ \psi_m = -(1 - m) \, u_{m + \frac{1}{2}} \,, \\
    \delta_- j &= g_- \,, \quad \delta_- j_+ = g_+ \,, \quad \delta_- \widetilde{g}_+ = -t \,, \\
    \delta_- w_m &= \psi_{m - \frac{1}{2}} \,, \quad \delta_- \chi_m = -u_{m - \frac{1}{2}} \,,
  \end{aligned}
\end{equation}
where coordinates with subscripts out of range are set to zero, and they annihilate all of the other coordinates; they are extended to $S_0$ as odd derivations, as in Sections \ref{sec:arcs} and \ref{sec:N3}, and $\delta_\pm$ shifts the $T$-degree by $\pm1/2$. We find three polynomials of conformal dimension five, $P^{(a)}$, where $a = -2, 0, 2$ is the $\mathfrak{su}(2)$ flavor charge, such that
\begin{equation}\label{eqn:N4SU3Jac}
    I_0 = \sum_{a = -2, 0, 2} \Big( \operatorname{Jac}(P^{(a)}) + \operatorname{Jac}(\delta_+ P^{(a)}) + \operatorname{Jac}(\delta_- P^{(a)}) \Big) \,,
\end{equation}
where $\operatorname{Jac}(f)$ is the Jacobian ideal defined in Section \ref{sec:arcs}; we have verified this equality exactly. The three superpotentials form the adjoint representation of the $\mathfrak{su}(2)$ flavor symmetry: $P^{(2)}$ and $P^{(-2)}$ are obtained from $P^{(0)}$ by acting with the raising and lowering operators of the $\mathfrak{su}(2)$ flavor transformations.

Since the superpotentials $P^{(\pm 2)}$ are determined by $P^{(0)}$ via the $\mathfrak{su}(2)$ flavor symmetry, it suffices to write down the superpotential of vanishing flavor charge, which is
\begin{equation}\label{eqn:N4SU3P0}
  \begin{aligned}
    P^{(0)} = \,\, & r_- r_+ + t^2 j + \frac{1}{2} t \left( g_+ \widetilde{g}_- + g_- \widetilde{g}_+ \right) \\
    & + t \left( -\frac{1}{6} j^3 + \frac{2}{3} j j_+ j_- + \frac{1}{4} w_{-3/2} w_{3/2} - \frac{1}{12} w_{-1/2} w_{1/2} \right) - \frac{1}{6} j^2 \left( g_+ \widetilde{g}_- + g_- \widetilde{g}_+ \right) \\
    & + \frac{1}{3} j \left( j_+ g_- \widetilde{g}_- + j_- g_+ \widetilde{g}_+ \right) - \frac{1}{12} j \left( \psi_{-1} \chi_1 + \psi_0 \chi_0 + \psi_1 \chi_{-1} \right) \\
    & + \frac{1}{12} j_+ \left( \psi_{-1} \chi_0 + \psi_0 \chi_{-1} \right) + \frac{1}{6} j_- \left( \psi_0 \chi_1 + \psi_1 \chi_0 \right) \\
    & + \frac{1}{12} g_+ \left( w_{-3/2} \chi_1 - w_{1/2} \chi_{-1} \right) - \frac{1}{12} g_- \left( w_{-1/2} \chi_1 - 3 w_{3/2} \chi_{-1} \right) \\
    & - \frac{1}{12} \widetilde{g}_+ \left( w_{-3/2} \psi_1 - w_{1/2} \psi_{-1} \right) + \frac{1}{12} \widetilde{g}_- \left( w_{-1/2} \psi_1 - 3 w_{3/2} \psi_{-1} \right) \\
    & - \frac{1}{36} \Big( 2 j^2 w_{-1/2} w_{1/2} - 2 j j_+ w_{-3/2} w_{1/2} - j j_+ w_{-1/2}^2 - 6 j j_- w_{-1/2} w_{3/2} \\
    & \qquad\quad - 4 j j_- w_{1/2}^2 + j_+^2 w_{-3/2} w_{-1/2} + 6 j_+ j_- w_{-3/2} w_{3/2} \\
    & \qquad\quad + 2 j_+ j_- w_{-1/2} w_{1/2} + 12 j_-^2 w_{1/2} w_{3/2} \Big) \,,
  \end{aligned}
\end{equation}
where the fermionic coordinates are ordered as written, and $r_\pm$ are the two minimal generators of $I_0$ at conformal dimension $5/2$ in equation \eqref{eqn:N4SU3r}. As for the superpotentials $P_p$ of Section \ref{sec:N3}, $P^{(0)}$ is not homogeneous in the $T$-grading: its terms have $T$-degrees three, four, and five. Apart from $r_- r_+$, it does not depend on the coordinates $u_{\pm 1/2}$, and differentiating with respect to $u_{\pm 1/2}$ reproduces $r_\mp$. This mirrors the $\mathfrak{su}(2)$ case: after the substitution described in Section \ref{sec:N3}, $P_2 = -s^2/4 + \cdots$, where $s = t - j_+ j_- - j^2$ is the Sugawara relation in equation \eqref{eqn:N4ideal}, and the remaining terms do not depend on $t$. There is, however, a difference: $P_2$ is invariant under the $\mathfrak{su}(2)$ flavor symmetry, whereas, for $\mathfrak{su}(3)$, no invariant superpotential of conformal dimension five generates $I_0$ in this way. In fact, a single superpotential of vanishing flavor charge cannot suffice: $I_0$ contains bosonic relations of conformal dimension three with flavor charge $\pm 2$, which cannot arise as derivatives of such a superpotential or of its two supersymmetry variations. We have also listed the three superpotentials, together with the action of $\delta_\pm$ on the coordinates, in the file {\tt N4SU3.m}.

\subsection{Superspace formulation}\label{sec:N4superspace}

In Sections \ref{sec:N4su2} and \ref{sec:N4su3}, the components of each $\mathcal{N}=4$ multiplet enter as independent coordinates, and the supersymmetry appears only through the supersymmetry variations. Here, we package the generators into superfields, which makes both the $\mathfrak{su}(2)$ flavor symmetry and the supersymmetry manifest, and we find that, for both gauge algebras, the ideal is generated by the components of a small number of superfield relations, and that it arises from the components of a single superpotential. We first introduce auxiliary coordinates, $z^+$ and $z^-$, which transform as a doublet of the $\mathfrak{su}(2)$ flavor symmetry, and package a multiplet of spin $j$, with components $O_m$ for $m = -j, \cdots, j$, as a homogeneous polynomial of degree $2j$ in $z^\pm$:
\begin{equation}\label{eqn:N4harmonic}
    O^{(j)} = \sum_{m=-j}^{j} O_m \, (z^+)^{j+m} (z^-)^{j-m} \,.
\end{equation}
The product $O^{(j_1)} O^{(j_2)}$ is then the component of spin $j_1 + j_2$ of the tensor product of the two multiplets. The other components are obtained by contracting with the invariant $\epsilon$-tensor, which leads to the $j$-product,
\begin{equation}\label{eqn:jproduct}
    \big(O^{(j_1)} O^{(j_2)}\big)_j = \sum_{i=0}^{r} (-1)^i \binom{r}{i} \left( \partial_{z^+}^{r-i} \partial_{z^-}^{i} O^{(j_1)} \right) \left( \partial_{z^+}^{i} \partial_{z^-}^{r-i} O^{(j_2)} \right) \,, \qquad r = j_1 + j_2 - j \,,
\end{equation}
for $|j_1 - j_2| \leq j \leq j_1 + j_2$; in particular, $\big(O^{(j_1)} O^{(j_2)}\big)_{j_1 + j_2} = O^{(j_1)} O^{(j_2)}$.

The four supercharges of the global $\mathcal{N}=4$ superconformal algebra form two doublets of the $\mathfrak{su}(2)$ flavor symmetry, and we introduce the corresponding Grassmann coordinates, $\theta_\pm$ and $\overline{\theta}_\pm$, which we contract with $z^\pm$ as $\theta = z^+ \theta_+ + z^- \theta_-$ and $\overline{\theta} = z^+ \overline{\theta}_+ + z^- \overline{\theta}_-$. For both gauge algebras, the multiplet with $p = 2$, which contains the small $\mathcal{N}=4$ superconformal algebra, is packaged into the superfield
\begin{equation}\label{eqn:N4J}
    \mathcal{J} = j^{(1)} + \theta g^{(1/2)} + \overline{\theta} \widetilde{g}^{(1/2)} + \theta \overline{\theta} \, t \,,
\end{equation}
where the multiplets are written as in equation \eqref{eqn:N4harmonic}, up to normalizations of the components which we specify below. We define the superderivatives, $\mathbb{D}$ and $\overline{\mathbb{D}}$, as the left derivatives with respect to $\theta$ and $\overline{\theta}$, acting on superfields such as $\mathcal{J}$,\footnote{Equivalently, $\mathbb{D} = \big( (z^+)^{-1} \partial_{\theta_+} + (z^-)^{-1} \partial_{\theta_-} \big) / 2$, and similarly for $\overline{\mathbb{D}}$.} so that
\begin{equation}
    \mathbb{D}\mathcal{J} = g^{(1/2)} + \overline{\theta} t \,, \qquad \overline{\mathbb{D}}\mathcal{J} = \widetilde{g}^{(1/2)} - \theta t \,, \qquad \overline{\mathbb{D}}\mathbb{D}\mathcal{J} = t \,.
\end{equation}
We then form products and $j$-products of superfields via equation \eqref{eqn:jproduct}, where the derivatives with respect to $z^\pm$ also act on $\theta$ and $\overline{\theta}$, and we refer to the coefficients of the monomials in $z^\pm$, $\theta_\pm$, and $\overline{\theta}_\pm$ as the components of the resulting superfield. Since the $j$-products act on $\theta$ and $\overline{\theta}$, the components can involve both $\theta_+$ and $\theta_-$.

For $\mathfrak{su}(2)$, the superfield $\mathcal{J}$ contains all of the coordinates of $S$, and, in terms of the coordinates of equation \eqref{eqn:N4ideal}, we take
\begin{equation}
  \begin{gathered}
    j^{(1)} = -\frac{1}{2} j_+ (z^+)^2 + j z^+ z^- + \frac{1}{2} j_- (z^-)^2 \,, \\
    g^{(1/2)} = g_+ z^+ + g_- z^- \,, \qquad \widetilde{g}^{(1/2)} = -\widetilde{g}_+ z^+ + \widetilde{g}_- z^- \,.
  \end{gathered}
\end{equation}
Then the ideal $I$ in equation \eqref{eqn:N4ideal} is generated by the components of a single superfield relation,
\begin{equation}\label{eqn:N4su2superrel}
    (\mathcal{J}\mathcal{J})_0 + 2 \, \overline{\mathbb{D}}\mathbb{D}\mathcal{J} = 0 \,,
\end{equation}
whose lowest component, obtained by setting $\theta_\pm = \overline{\theta}_\pm = 0$, is $2(t - j_+ j_- - j^2)$, that is, the Sugawara relation; the higher components give the remaining generators of $I$. Furthermore, consider the superfield of spin zero and conformal dimension four,
\begin{equation}\label{eqn:N4su2superpot}
    \mathcal{P} = \big( \overline{\mathbb{D}}\mathbb{D}\mathcal{J} \big)^2 + (\mathcal{J}\mathcal{J})_0 \, \overline{\mathbb{D}}\mathbb{D}\mathcal{J} + \big( \mathcal{J} (\mathbb{D}\mathcal{J} \, \overline{\mathbb{D}}\mathcal{J})_1 \big)_0 \,.
\end{equation}
We find that $I = \sum_c \operatorname{Jac}(\mathcal{P}_c)$, where the sum runs over all of the components, $\mathcal{P}_c$, of $\mathcal{P}$, and where $\operatorname{Jac}(f)$ is the Jacobian ideal defined in Section \ref{sec:arcs}; we have verified this equality exactly. The superfield relation in equation \eqref{eqn:N4su2superrel} is obtained by formally differentiating $\mathcal{P}$ with respect to $\overline{\mathbb{D}}\mathbb{D}\mathcal{J}$, treated as an independent superfield. Moreover, after the substitution in Section \ref{sec:N3}, the lowest component of $\mathcal{P}$ is exactly $-4P_2$, where $P_2$ is the superpotential of the rank-one $\mathcal{N}=3$ SCFT with $p = 2$; thus, $\mathcal{P}$ is the superspace counterpart of $P_2$, which is invariant under the $\mathfrak{su}(2)$ flavor symmetry.

For $\mathfrak{su}(3)$, the generators lie in two short multiplets of the global supersymmetry, the multiplets with $p = 2$ and $p = 3$ in Table \ref{tbl:N4SU3SG}, and we package the latter into the superfield:\footnote{For other gauge algebras, the strong generators do not, in general, all lie in short multiplets; for example, for gauge algebras of type $D_n$, we expect, from the index and from the free-field realizations of \cite{Bonetti:2018fqz}, that there is a strong generator in a long multiplet.}
\begin{equation}\label{eqn:N4W}
    \mathcal{W} = w^{(3/2)} + \theta \psi^{(1)} + \overline{\theta} \chi^{(1)} + \theta \overline{\theta} u^{(1/2)} \,.
\end{equation}
In terms of the coordinates of Table \ref{tbl:N4SU3SG}, whose normalizations differ from those of the $\mathfrak{su}(2)$ case, all of the components of $\mathcal{J}$ and $\mathcal{W}$ are as in equation \eqref{eqn:N4harmonic}, with the exception of four components of $\mathcal{W}$, whose normalizations we adjust as
\begin{equation}
    w^{(3/2)} = w_{3/2} (z^+)^3 + w_{1/2} (z^+)^2 z^- + \frac{1}{2} w_{-1/2} z^+ (z^-)^2 + \frac{1}{6} w_{-3/2} (z^-)^3 \,,
\end{equation}
and similarly the coefficients of $(z^-)^2$ in $\psi^{(1)}$ and $\chi^{(1)}$ are $\psi_{-1}/2$ and $\chi_{-1}/2$. The superpotential is now a superfield of spin one and conformal dimension five:
\begin{equation}\label{eqn:N4superpotential}
  \begin{aligned}
    \mathcal{P} = \,\, & \frac{1}{2} \big( \overline{\mathbb{D}}\mathbb{D}\mathcal{W} \big)^2 + \frac{1}{6} \big( \overline{\mathbb{D}}\mathbb{D}\mathcal{W} \big) (\mathcal{J}\mathcal{W})_{1/2} + \frac{1}{96} \big( (\mathcal{J}\mathcal{W})_{1/2} \big)^2 \\
    & + \mathcal{J} \big( \overline{\mathbb{D}}\mathbb{D}\mathcal{J} \big)^2 + \frac{1}{2} \big( \overline{\mathbb{D}}\mathbb{D}\mathcal{J} \big) (\mathbb{D}\mathcal{J} \, \overline{\mathbb{D}}\mathcal{J})_1 + \frac{1}{12} (\mathcal{J}\mathcal{J})_0 \, \mathcal{J} \, \overline{\mathbb{D}}\mathbb{D}\mathcal{J} \\
    & + \frac{1}{48} \big( \overline{\mathbb{D}}\mathbb{D}\mathcal{J} \big) (\mathcal{W}\mathcal{W})_1 + \frac{1}{12} \mathcal{J} \big( \mathcal{J} (\mathbb{D}\mathcal{J} \, \overline{\mathbb{D}}\mathcal{J})_1 \big)_0 + \frac{1}{72} \big( \mathcal{J} (\mathbb{D}\mathcal{W} \, \overline{\mathbb{D}}\mathcal{W})_2 \big)_1 \\
    & - \frac{1}{72} \mathcal{J} (\mathbb{D}\mathcal{W} \, \overline{\mathbb{D}}\mathcal{W})_0 + \frac{1}{24} \big( (\overline{\mathbb{D}}\mathcal{J} \, \mathcal{W})_1 \mathbb{D}\mathcal{W} \big)_1 - \frac{1}{24} \big( (\mathbb{D}\mathcal{J} \, \mathcal{W})_1 \overline{\mathbb{D}}\mathcal{W} \big)_1 \,.
  \end{aligned}
\end{equation}
Its lowest component is exactly the triplet of superpotentials in Section \ref{sec:N4su3}:
\begin{equation}
    \mathcal{P} \big|_{\theta = \overline{\theta} = 0} = -\frac{1}{2} P^{(2)} (z^+)^2 + P^{(0)} z^+ z^- + \frac{1}{2} P^{(-2)} (z^-)^2 \,,
\end{equation}
with $P^{(0)}$ as in equation \eqref{eqn:N4SU3P0}, and with $P^{(\pm 2)}$ as listed in {\tt N4SU3.m}. Furthermore, the components with a single $\theta_\pm$, or a single $\overline{\theta}_\pm$, and no other Grassmann coordinates are supersymmetry variations of the $P^{(a)}$, associated with the supercurrents $G_\pm$, or $\widetilde{G}_\pm$, respectively. As for $\mathfrak{su}(2)$, we find that
\begin{equation}\label{eqn:N4superJac}
    I_0 = \sum_c \operatorname{Jac}(\mathcal{P}_c) \,,
\end{equation}
where the sum runs over all of the components of $\mathcal{P}$, including those involving more than one of the $\theta_\pm$ and $\overline{\theta}_\pm$; we have verified this equality exactly. Thus, the superfield $\mathcal{P}$ combines the three superpotentials and their supersymmetry variations in equation \eqref{eqn:N4SU3Jac} into a single object. The difference between the two gauge algebras that we observed in Section \ref{sec:N4su3} is now transparent: for $\mathfrak{su}(2)$, the superpotential has spin zero, whereas, for $\mathfrak{su}(3)$, it has spin one.

The superfield $\mathcal{P}$ also determines a compact set of generators of $I_0$. Formally differentiating $\mathcal{P}$ with respect to $\overline{\mathbb{D}}\mathbb{D}\mathcal{W}$ and $\overline{\mathbb{D}}\mathbb{D}\mathcal{J}$, treated as independent superfields, gives, up to normalization, the first two of the relations
\begin{equation}\label{eqn:N4superrels}
  \begin{aligned}
    & (\mathcal{J}\mathcal{W})_{1/2} + 6 \, \overline{\mathbb{D}}\mathbb{D}\mathcal{W} = 0 \,, \\
    & (\mathcal{W}\mathcal{W})_1 + 24 \, (\mathbb{D}\mathcal{J} \, \overline{\mathbb{D}}\mathcal{J})_1 + 96 \, \mathcal{J} \, \overline{\mathbb{D}}\mathbb{D}\mathcal{J} + 4 \, (\mathcal{J}\mathcal{J})_0 \, \mathcal{J} = 0 \,, \\
    & 24 \, \big( \overline{\mathbb{D}}\mathbb{D}\mathcal{J} \big)^2 - (\mathbb{D}\mathcal{W} \, \overline{\mathbb{D}}\mathcal{W})_0 + 2 \, (\mathcal{J}\mathcal{J})_0 \, \overline{\mathbb{D}}\mathbb{D}\mathcal{J} = 0 \,,
  \end{aligned}
\end{equation}
and we find that the components of these three superfields generate $I_0$. The lowest component of the first relation is $6(r_+ z^+ + r_- z^-)$, with $r_\pm$ as in equation \eqref{eqn:N4SU3r}, and the lowest component of the second relation is proportional to the three minimal generators of $I_0$ of conformal dimension three that depend on $t$, which are the analogues of the Sugawara relation of the $\mathfrak{su}(2)$ theory. The first two relations alone do not suffice: they miss a single bosonic relation of conformal dimension four with vanishing flavor charge, which is supplied by the third relation, and which is determined only up to the ideal generated by the first two. As for the superpotentials, the relations in equations \eqref{eqn:N4su2superrel} and \eqref{eqn:N4superrels} are not homogeneous in the $T$-grading; for example, the two terms in the first relation in equation \eqref{eqn:N4superrels} have $T$-degrees $5/2$ and $3/2$.

\section{Rank-one \texorpdfstring{\boldmath{$\mathcal{N}=3$}}{N=3} SCFTs}\label{sec:N3}

Another class of 4d SCFTs where the chiral algebra has fermionic strong generators is the theories with enhanced $\mathcal{N}=3$ supersymmetry. We consider the rank-one theories \cite{Garcia-Etxebarria:2015wns,Nishinaka:2016hbw,Aharony:2016kai}, which are labelled by $\mathbb{Z}_p$, for $p = 2, 3, 4, 6$, and whose associated VOAs are given in \cite{Nishinaka:2016hbw,Bonetti:2018fqz}. The $p = 2$ case has enhanced $\mathcal{N}=4$ supersymmetry, and its VOA is that of $\mathcal{N}=4$ super Yang--Mills with gauge algebra $\mathfrak{su}(2)$, which we studied in Section \ref{sec:N4}. We will find that, for each $p$, the ideal is generated by the derivatives of a single superpotential and of its two supersymmetry variations, and that, for $p = 2$, this construction reproduces exactly the ideal in equation \eqref{eqn:N4ideal}.

In each case, the VOA admits a free-field realization \cite{Bonetti:2018fqz}, and it is strongly generated by the operators listed in Table \ref{tbl:N3SG}: the $\mathcal{N}=2$ stress tensor multiplet, $(J, G, \widetilde{G}, T)$, together with $(W, G_W)$ and $(\overline{W}, \widetilde{G}_{\overline{W}})$. For $p = 2$, the stress tensor is the Sugawara stress tensor, and there are only seven strong generators; we nevertheless include $T$ for uniformity. As for the free vector multiplet and for $\mathcal{N}=4$ super Yang--Mills, the $T$-degrees are fixed by the 4d charges of the corresponding Schur operators. This motivates us to consider the $\mathbb{Z}_2$-graded-commutative ring
\begin{equation}\label{eqn:N3S}
    S^{(p)} = \mathbb{C}[j, t, w, \overline{w}] \otimes \Lambda[g, \widetilde{g}, \eta, \overline{\eta}] \,,
\end{equation}
where the coordinates correspond to the strong generators of the VOA as
\begin{equation}
    (j, t, w, \overline{w}, g, \widetilde{g}, \eta, \overline{\eta}) \leftrightarrow (J, T, W, \overline{W}, G,\widetilde{G}, G_W,\widetilde{G}_{\overline{W}}) \,,
\end{equation}
and the $q$- and $T$-degrees of the coordinates are as in Table \ref{tbl:N3SG}.

\begin{table}[H]
    \centering
    \begin{threeparttable}
    \begin{tabular}{ccccccccc}
        \toprule
        Strong Gen. & $J$ & $T$ & $G$ & $\widetilde{G}$ & $W$ & $\overline{W}$ & $G_W$ & $\widetilde{G}_{\overline{W}}$ \\\midrule
        $F$ & $0$ & $0$ & $1$ & $1$ & $0$ & $0$ & $1$ & $1$ \\[1pt]
        $h$ & $1$ & $2$ & $\frac{3}{2}$ & $\frac{3}{2}$ & $\frac{p}{2}$ & $\frac{p}{2}$ & $\frac{p+1}{2}$  & $\frac{p+1}{2}$ \\[3pt]
        $T$ & $1$ & $1$ & $\frac{3}{2}$ & $\frac{1}{2}$ & $\frac{p}{2}$ & $\frac{p}{2}$ & $\frac{p+1}{2}$ & $\frac{p-1}{2}$ \\\bottomrule
    \end{tabular}
    \end{threeparttable}
    \caption{The strong generators of the chiral algebra associated to the rank-one $\mathbb{Z}_p$ $\mathcal{N}=3$ SCFTs. We list the fermion number ($F$), conformal dimension ($h$), and $T$-grading $(T)$. There is also an additional $\mathfrak{u}(1)$ flavor symmetry, which we ignore here for simplicity. For $p = 2$, these reproduce the strong generators in Table \ref{tbl:N4SG}, together with the stress tensor, which is then not a strong generator.}
    \label{tbl:N3SG}
\end{table}

The null relations of these VOAs were computed explicitly using the free-field realization in terms of $\beta\gamma bc$ ghost systems in \cite{Bonetti:2018fqz} (see also \cite{Agarwal:2021oyl}). From them, we can determine the ideals, $I_p$, such that $S^{(p)} \, / \, I_p$ is Zhu's $C_2$-algebra. In fact, we find that each of these ideals can be derived from a Jacobian-like construction, which we now explain. We begin with the following polynomial in $S^{(p)}$, which we refer to as the superpotential:
\begin{equation}
    P_p = t(A_p(j,t) - p^p w \overline{w}) + B_p(j,t)g\widetilde{g} - p^{p-2} j\eta\overline{\eta} + p^{p-1} (g w \overline{\eta} + \widetilde{g} \overline{w} \eta ) \,.
\end{equation}
The auxiliary polynomials in $j$ and $t$ can be written as
\begin{equation}
    \begin{aligned}
        A_p &= j^p
-\frac{(p-1)p!}{2(p-2)!}j^{p-2}t
+\frac{2(p-3)p!}{3(p-4)!}j^{p-4}t^2
-\frac{3(p-5)p!}{4(p-6)!}j^{p-6}t^3 \,, \\
        B_p &= j^{p-1}
-\frac{p!}{(p-3)!}j^{p-3}t
+\frac{3p!}{2(p-5)!}j^{p-5}t^2  \,,
    \end{aligned}
\end{equation}
where the terms for which the argument of the factorial in the denominator becomes negative are assumed to be zero.\footnote{We can write $A_p$ and $B_p$ in terms of a differential operator, $\mathcal{D} = t \partial_j^2$, as
\begin{equation*}
    A_p = \left[ 1 + (j \partial_j + 1)\left(\frac{1}{1 + \mathcal{D}} - \frac{\log(1 + \mathcal{D})}{\mathcal{D}}\right)\right] j^p \,, \qquad B_p = \left[ 1 + p\left(\log(1 + \mathcal{D}) - \frac{2\mathcal{D}}{1 + \mathcal{D}}\right)\right] j^{p-1} \,. 
\end{equation*}
The physical significance of these formulae remains to be understood.} The superpotential is homogeneous of $q$-degree $p + 2$, but it is not homogeneous in the $T$-grading, since $j^2$ and $t$ have the same $q$-degree but different $T$-degrees. The superfield expansions
\begin{equation}\label{eqn:N3superfields}
    \mathcal{T} = J + \theta^+ G + \theta^- \widetilde{G} + \theta^+ \theta^- T \,, \qquad \mathcal{W} = W + \theta^+ G_W \,, \qquad \overline{\mathcal{W}} = \overline{W} - \theta^- \widetilde{G}_{\overline{W}} \,,
\end{equation}
define two fermionic supersymmetry variations, $\delta_+$ and $\delta_-$, which act on the coordinates as
\begin{equation}
  \begin{aligned}
    \delta_+ j &= g \,, \qquad \delta_+ \widetilde{g} = - t \,, \qquad \delta_+ w = \eta \,, \\
    \delta_- j &= \widetilde{g} \,, \qquad \delta_- g = t \,, \qquad \delta_- \overline{w} = -\overline{\eta} \,,
  \end{aligned}
\end{equation}
and annihilate all other coordinates; they are extended to $S^{(p)}$ as odd derivations, as in Section \ref{sec:arcs}. The variations raise the $q$-degree by $1/2$, and $\delta_\pm$ shifts the $T$-degree by $\pm 1/2$; thus, the $T$-degrees of all of the coordinates in Table \ref{tbl:N3SG} are determined by those of $j$, $w$, and $\overline{w}$. With these two variations and $P_p$ in hand, we can construct two new polynomials:
\begin{equation}\label{eqn:superpartnerpotentials}
    \begin{aligned}
        \delta_+ P_p &= t(\partial_j A_p + B_p)g - p^{p-1} (p+1) t \overline{w} \eta - p^{p-2} (p+1) g \eta \overline{\eta} \,, \\
        \delta_- P_p &= t(\partial_j A_p + B_p)\widetilde{g} + p^{p-1} (p+1) t w \overline{\eta} - p^{p-2} (p+1) \widetilde{g} \eta \overline{\eta} \,.
    \end{aligned}
\end{equation}
Then, we consider the following sum of Jacobian ideals in $S^{(p)}$:
\begin{equation}\label{eqn:dora}
    I_p = \operatorname{Jac}(P_p) + \operatorname{Jac}(\delta_+ P_p) + \operatorname{Jac}(\delta_- P_p) \,,
\end{equation}
where $\operatorname{Jac}(f)$ is the Jacobian ideal defined in Section \ref{sec:arcs}. It can be explicitly verified that this ideal is identical to the ideal worked out from the free-field realizations of the VOAs. Therefore, equation \eqref{eqn:dora} is a concise, but somewhat redundant, presentation of the filtered graded ring
\begin{equation}
    R_{\mathbb{Z}_p} = S^{(p)} \, / \, I_p \,,
\end{equation}
which we propose underlies both the Higgs branch and the Macdonald index of the rank-one $\mathcal{N}=3$ SCFTs, for $p = 2, 3, 4, 6$. Since the fermionic coordinates are nilpotent, the reduced ring is determined by the purely bosonic part of $I_p$, and we find
\begin{equation}
    \operatorname{Spec}(R_{\mathbb{Z}_p})_\text{red} = \operatorname{Spec}\left( \mathbb{C}[j, t, w, \overline{w}] \, / \, (t, j^p - p^p w \overline{w}) \right) \cong \mathbb{C}^2 / \mathbb{Z}_p \,,
\end{equation}
which is the Higgs branch of the rank-one $\mathbb{Z}_p$ theory.

For $p = 2$, we have $A_2 = j^2 - t$ and $B_2 = j$, and the generator of $I_2$ obtained by differentiating the superpotential with respect to $t$,
\begin{equation}
    \partial_t P_2 = j^2 - 2t - 4 w \overline{w} \,,
\end{equation}
is linear in $t$: it expresses the stress tensor as the Sugawara stress tensor, which is why $T$ is not a strong generator in this case. For $p \geq 3$, $\partial_t P_p$ has no term proportional to $t$ alone, and so it cannot be used to eliminate $t$; correspondingly, $T$ is a strong generator. In fact, the whole of $I_2$ coincides with the ideal of $\mathcal{N}=4$ super Yang--Mills with gauge algebra $\mathfrak{su}(2)$: under the substitution
\begin{equation}
    (j, t, w, \overline{w}, g, \widetilde{g}, \eta, \overline{\eta}) \,\rightarrow\, \left(j, \tfrac{1}{2} t, \tfrac{1}{2} j_-, -\tfrac{1}{2} j_+, \widetilde{g}_+, \tfrac{1}{2} g_-, -\widetilde{g}_-, -\tfrac{1}{2} g_+\right) \,,
\end{equation}
where the coordinates on the right-hand side are those of Section \ref{sec:N4su2}, the ideal $I_2$ becomes exactly the ideal $I$ in equation \eqref{eqn:N4ideal}. Thus, the superpotential $P_2$ accounts for both the null relations and the Sugawara relation of Section \ref{sec:N4}, including the failure of the latter to be homogeneous in the $T$-grading.

The authors of \cite{Agarwal:2021oyl} have provided a {\tt Mathematica} notebook which computes the Macdonald index of the rank-one theories from the free-field realization of the VOA, by subtracting the null states level by level, and they have used it to determine the Macdonald index up to order $q^7$, and its Schur limit up to order $q^8$. In principle, this can be carried out to any order, but in practice the computation becomes extremely slow at higher orders; running their notebook, we have extended the Macdonald index up to order $q^{10}$. We have computed the Hilbert series of the arc space up to and including order $q^{22}$, $q^{23}$, and $q^{24}$ for $p = 3$, $4$, and $6$, respectively, and we have verified that it agrees with the Macdonald index up to and including order $q^{10}$, that is, for the first twenty-one coefficients of the expansion in powers of $q^{1/2}$, each of which is a polynomial in $T^{1/2}$. We leave the higher-order terms as a prediction for the Macdonald index. For $p = 3$, there is an independent check of the higher-order terms in the Schur limit. The Schur index of the $\mathbb{Z}_3$ theory can be computed from the $\mathcal{N}=1$ Lagrangian description of \cite{Zafrir:2020epd}, and it has recently been obtained in closed form \cite{Guo:2026lsl}:
\begin{equation}
    I_\text{Schur}^{\mathbb{Z}_3}(q) = \frac{q^{-5/8}}{3} \left( \frac{\eta(q)^4}{\eta(q^{1/2})^2} - \frac{\eta(q^{1/2}) \, \eta(q^3)^2}{\eta(q)^2 \, \eta(q^{3/2})} \right) = 1 + q + q^2 + 2q^3 - 2q^{7/2} + 3q^4 + \cdots \,,
\end{equation}
where $\eta(q) = q^{1/24} \prod_{n \geq 1} (1 - q^n)$, and we have removed the overall factor $q^{-c/24} = q^{5/8}$, with $c = -15$. The Schur limit of the Hilbert series agrees with this expression up to and including order $q^{22}$. Since $S^{(p)} \, / \, I_p$ is Zhu's $C_2$-algebra, this agreement also indicates that these VOAs are classically free, as we found for $p = 2$ in Section \ref{sec:N4}. The Hilbert series for all $p$ are summarized in the {\tt Mathematica} notebook {\tt ArcSpaceHS.nb} attached to the arXiv submission of this paper.

\section{The \texorpdfstring{\boldmath{$\widehat{E}_6(SU(2))$}}{E6(SU(2))} SCFT and the \texorpdfstring{\boldmath{$\mathcal{A}(6)$}}{A(6)} VOA}\label{sec:A6}

In each of the examples so far, the arc space of Zhu's $C_2$-algebra has reproduced the Schur index to the order that we have checked, which indicates that the VOA is classically free. We now turn to an example where this fails, and where the generalized arc space $J_{\infty,1}(S, I_1)$ is necessary. The 4d $\mathcal{N}=2$ SCFT in question is variously known as $(A_2, D_4)$, $\widehat{E}_6(SU(2))$, $\mathcal{T}_{(3,2)}$, amongst others \cite{Kang:2021lic,Buican:2016arp,Buican:2020moo,DelZotto:2015rca}, and it is convincingly argued \cite{Buican:2016arp} that the associated chiral algebra is the $\mathcal{A}(6)$ VOA of Feigin--Feigin--Tipunin \cite{Feigin:2007sp,MR2766985}.\footnote{In the context of the SCFT/VOA correspondence, the $\mathcal{A}(6)$ chiral algebra has been well-studied, for example, \cite{Buican:2016arp,Buican:2020moo,Kang:2021lic,Pan:2025gzh,Jiang:2024baj,Jiang:2026ffv}.}

The strong generators of this VOA are the (bosonic) stress tensor, $T$, of conformal dimension two, and two fermionic operators, $\Psi$ and $\widetilde{\Psi}$, both of conformal dimension four. In particular, $\mathcal{A}(6)$ is the extension of the Virasoro algebra with $c = -24$ by the Virasoro primaries $\Psi$ and $\widetilde{\Psi}$, which satisfy the OPEs \cite{Feigin:2007sp,Jiang:2024baj}
\begin{equation}
\label{eqn:A6TOPE}
\begin{gathered}
T(z)\Psi(0) \sim \frac{4\Psi}{z^2} + \frac{\Psi'}{z} \,, \qquad T(z) \widetilde{\Psi}(0) \sim \frac{4\widetilde{\Psi}}{z^2} + \frac{\widetilde{\Psi}'}{z} \,,
\end{gathered}
\end{equation}
together with
\begin{equation}\label{eqn:A6PsiPsitOPE}
\begin{aligned}
\Psi(z)\widetilde{\Psi}(0) \sim{} & -\frac{6}{z^8} + \frac{2T}{z^6} + \frac{T'}{z^5} + \frac{3\left(T'' - T^2\right)}{7z^4} + \frac{2T^{(3)} - 9T'T}{21z^3} \\
& + \frac{-48\left(T'\right)^2 - 84 T''T + 36T^3 + 7T^{(4)}}{420z^2} \\ & + \frac{60\left(-5T''T' + 6T'T^2 - 2T^{(3)}T\right) + 7T^{(5)}}{2800z} \,.
\end{aligned}
\end{equation}
This VOA also admits a free-field realization in terms of a single chiral boson, $\varphi$, satisfying $\varphi(z)\varphi(0) \sim \log z$. Explicitly, one has \cite{Feigin:2007sp,Jiang:2024baj}
\begin{equation}\label{eqn:A6freefield}
    \Psi(w) = e^{-\sqrt{3}\varphi(w)} \,, \qquad T = \frac{1}{2}\left(\partial \varphi\right)^2 + \frac{5}{2\sqrt{3}} \partial^2 \varphi \,,
\end{equation}
while $\widetilde{\Psi}$ is obtained from $\Psi$ via
\begin{equation}\label{eqn:A6freefield2}
    \widetilde{\Psi}(w) = \frac{1}{2\pi i} \oint dz \, e^{2\sqrt{3}\varphi(z)} \Psi(w) = P_5\left(\partial\varphi, \partial^2 \varphi, \dots, \partial^5\varphi\right) e^{\sqrt{3}\varphi(w)} \,,
\end{equation}
where $P_5$ is a (somewhat unilluminating) weight-five polynomial in the derivatives $\partial^j \varphi$, whose explicit form is straightforward to determine but which we do not reproduce here.

The vacuum character of the $\mathcal{A}(6)$ VOA is given via the following expression \cite{Feigin:2007sp,MR2766985,Buican:2016arp}:
\begin{equation}\label{eqn:A6character}
\chi_{\mathcal{A}(6)}(q) = \frac{1}{(q;q)_\infty}
\sum_{n=1}^{\infty} (-1)^{n-1} n
\left(
q^{\frac{3n^2 - n - 2}{2}} - q^{\frac{3n^2 + n - 2}{2}}
\right) \,.
\end{equation}
Furthermore, the $\widehat{E}_6(SU(2))$ theory can be obtained via the diagonal $\mathfrak{su}(2)$ gauging of the flavor symmetry of a triplet of $\mathcal{D}_3(SU(2))$ Argyres--Douglas theories. This provides us with a mechanism to compute the Macdonald index of $\widehat{E}_6(SU(2))$, or, equivalently, the refined character of the $\mathcal{A}(6)$ chiral algebra. The Macdonald index of $\mathcal{D}_3(SU(2))$, where $z$ is the fugacity for the $\mathfrak{su}(2)$ flavor symmetry, can be determined from the TQFT expression of \cite{Song:2015wta}, and it can be written as (see also \cite{Buican:2015tda, Andrews:2025krn, Andrews:2025tko, Kim:2025klh} for alternative expressions)
\begin{equation}\label{eqn:D3SU2Mac}
  \begin{aligned}
    I_\text{Mac}^{\mathcal{D}_3(SU(2))}(q, T, z) &= \frac{1}{(qT; q)_\infty^2 (qTz^2; q)_\infty (qTz^{-2}; q)_\infty} \sum_{\mu = 0}^\infty (-1)^\mu q^{\frac{3\mu(\mu+1)}{2}} T^{2\mu} (1 - q^{2\mu+1} T) \\
    &\qquad \times (q^{2\mu+2} T^2; q)_\infty \frac{(qT; q)_\mu}{(q; q)_\mu} \sum_{m=0}^{2\mu} \begin{bmatrix} 2\mu \\ m \end{bmatrix}_q (qT; q)_m (qT; q)_{2\mu - m} \, z^{2(m - \mu)} \,,
  \end{aligned}
\end{equation}
where $(a; q)_n$ is the $q$-Pochhammer symbol defined in Section \ref{sec:vec}, and the $q$-binomial coefficient is
\begin{equation}
    \begin{bmatrix} n \\ m \end{bmatrix}_q = \frac{(q; q)_n}{(q; q)_m (q; q)_{n-m}} \,.
\end{equation}
The Macdonald index of $\widehat{E}_6(SU(2))$ is then
\begin{equation}\label{eqn:refA6gauge}
    I_\text{Mac}^{\widehat{E}_6(SU(2))}(q, T) = \int [dz] I_\text{Mac}^{\text{vec}}(q, T, z) \left(I_\text{Mac}^{\mathcal{D}_3(SU(2))}(q, T, z)\right)^3 \,.
\end{equation}

Inspired by the strong generators of $\mathcal{A}(6)$, we consider the $\mathbb{Z}_2$-graded-commutative ring with one bosonic and two fermionic generators:
\begin{equation}\label{eqn:A6S}
    S = \mathbb{C}[x] \otimes \Lambda[y, z] \,,
\end{equation}
where the coordinates correspond to the strong generators as $(x, y, z) \leftrightarrow (T, \Psi, \widetilde{\Psi})$, and where the $q$- and $T$-degrees are
\begin{equation}\label{eqn:A6gradings}
    \operatorname{deg}(x) = (2,1) \,, \quad \operatorname{deg}(y) = (4, 3) \,, \quad \operatorname{deg}(z) = (4,2) \,.
\end{equation}
As in the previous sections, the stress tensor has $T$-degree one. The $T$-degrees of $y$ and $z$ can be read off from the Macdonald index: the coefficient of $q^4$ in equation \eqref{eqn:refA6gauge} is $T - T^3$, which receives contributions from $x_2$ and $x_0^2$, with $T$-degrees one and two, and from $y_0$ and $z_0$, which must therefore have $T$-degrees three and two. The null relations of $\mathcal{A}(6)$ that survive in Zhu's $C_2$-algebra are $T\Psi$ and $T\widetilde{\Psi}$, at conformal dimension six, and $3T^4 - 140\Psi\widetilde{\Psi}$, at conformal dimension eight, and so
\begin{equation}\label{eqn:A6I0}
    R_\mathbb{V} = S \, / \, I_0 \,, \qquad I_0 = (xy, xz, 3x^4 - 140yz) \,.
\end{equation}
The last relation is not homogeneous in the $T$-grading, since $x^4$ and $yz$ have $T$-degrees four and five, respectively, and thus $R_\mathbb{V}$ is a filtered graded ring. Since $x^4$ is proportional to the nilpotent element $yz$, the coordinate $x$ is nilpotent, and $\operatorname{Spec}(R_\mathbb{V})_\text{red}$ is a point, consistent with the trivial Higgs branch of the $\widehat{E}_6(SU(2))$ SCFT.

We now compare the Hilbert series of the arc space of $R_\mathbb{V}$ with the vacuum character in equation \eqref{eqn:A6character}. We find that
\begin{equation}
    \operatorname{HS}_{q,q,-1}(J_\infty(R_\mathbb{V})) - \chi_{\mathcal{A}(6)}(q) = 3q^9 + 3q^{10} + 3q^{11} + \cdots \,,
\end{equation}
so the two agree up to conformal dimension eight, but not at conformal dimension nine. Since the map $\Phi$ in equation \eqref{eqn:CF} is surjective, and preserves the conformal dimension and the fermion number, it cannot be injective, and so $\mathcal{A}(6)$ is not classically free. Thus, as anticipated in Section \ref{Intro}, $\mathcal{A}(6)$ provides a counterexample to the conjecture of \cite{RastelliTalk1,RastelliTalk2}.

To determine the missing relations, we use the free-field realization in equations \eqref{eqn:A6freefield} and \eqref{eqn:A6freefield2}. At a given conformal dimension, we map the normally ordered monomials in $T$, $\Psi$, $\widetilde{\Psi}$, and their derivatives, which correspond to the monomials of $S_\infty$, into the Fock module of $\varphi$, and we determine the linear relations among them. For each such relation, the component with the lowest number of derivatives is an element of the kernel, $K$, of the map $S_\infty \rightarrow \operatorname{gr}_{\operatorname{Li}}(\mathbb{V})$ described in Section \ref{sec:arcs}. At conformal dimension nine, we find the following three elements of $K$ that do not belong to $DI_0$:
\begin{equation}\label{eqn:A6kernel}
    y_0 y_1 \,, \qquad z_0 z_1 \,, \qquad 9x_0^3x_1 + 70y_1z_0 - 280y_0z_1 \,.
\end{equation}
The first of these can be seen directly: from equation \eqref{eqn:A6freefield}, we have $\Psi(z) \Psi(0) = z^3 :\! e^{-\sqrt{3}\varphi(z)} e^{-\sqrt{3}\varphi(0)} \!:$, and therefore the normally ordered product $:\!\partial\Psi \Psi\!:$ vanishes identically in $\mathcal{A}(6)$, even though $y_0 y_1$ is non-zero in $J_\infty(R_\mathbb{V})$. The third element in equation \eqref{eqn:A6kernel} differs from $-175(y_0z_1 - y_1z_0)$ by a multiple of $D(3x_0^4 - 140y_0z_0)$, which belongs to $DI_0$. We therefore consider the ideal of $S_1$
\begin{equation}\label{eqn:A6I1}
  I_1 = D^{\leq 1} I_0 + (y_0y_1, z_0z_1, y_0z_1 - y_1z_0) \,,
\end{equation}
and we conjecture that $K = DI_1$, that is, that the kernel is generated, as a differential ideal, by relations of order at most one. Equivalently, $\operatorname{gr}_{\operatorname{Li}}(\mathbb{V}) \cong J_{\infty,1}(S, I_1)$.

We can now compute the Hilbert series of the generalized arc space. We begin with the Schur index, where we forget about the $T$-filtration, and we find
\begin{equation} 
  \begin{aligned}
    \operatorname{HS}_{q,q,-1}(J_{\infty,1}(S, I_1)) &= 1 + q^2 + q^3 + 2 q^6 + q^8 + q^{11} + 2 q^{12} + q^{15} + 2 q^{18}  \\&\qquad\qquad+
 2 q^{20} + q^{24} + q^{26} + 2 q^{27} + 2 q^{30} + q^{35} + 2 q^{36} \\&\qquad\qquad+ 2 q^{38} + 2 q^{42} + q^{47} + 3 q^{48} + 2 q^{51} + \mathcal{O}(q^{52}) \,.
  \end{aligned}
\end{equation}
This expression agrees with the vacuum character of the $\mathcal{A}(6)$ VOA, with the fermion number set to $-1$, as given in equation \eqref{eqn:A6character}, up to and including order $q^{51}$. We can also determine the Hilbert series keeping track of the $T$-filtration, in which case we must pass to the associated graded ring, since $I_1$ is not homogeneous in the $T$-grading. We find
\begin{equation}
  \begin{aligned}
    &\operatorname{HS}_{q,q,T,-1}(\operatorname{gr}(J_{\infty,1}(S, I_1))) = 1 + T q^2 + 
 T q^3 + (T - T^3) q^4 + (T - T^3) q^5 \\&\qquad + (T + T^2) q^6 + (T + T^2 - 
    T^3 - T^4) q^7 + (T + 2 T^2 - T^3 - T^4) q^8 \\&\qquad+ (T + 
    2 T^2 - T^3 - 2 T^4 ) q^9 + (T + 3 T^2 - T^3 - 3 T^4 ) q^{10} \\&\qquad+ (T + 3 T^2 - T^3 - 3 T^4 + T^6 ) q^{11} + (T + 4 T^2 - 4 T^4 + T^6 ) q^{12} \\&\qquad+ (T + 4 T^2 - 5 T^4 - T^5 + T^6) q^{13} + (T + 5 T^2 + T^3 - 6 T^4 - 2 T^5 + T^6) q^{14} &\\&\qquad+ (T + 5 T^2 + 2T^3 - 7T^4 - 3T^5 + 2T^6 + T^7)q^{15} + \mathcal{O}(q^{16}) \,.
  \end{aligned}
\end{equation}
This matches with the Macdonald index of the $\widehat{E}_6(SU(2))$ SCFT, as worked out from the gauging formula in equation \eqref{eqn:refA6gauge}. In fact, we have computed the Hilbert series of $\operatorname{gr}(J_{\infty,1}(S, I_1))$ up to and including order $q^{51}$, and it agrees with the series expansion of equation \eqref{eqn:refA6gauge} at every order, that is, for the first fifty-two coefficients of the expansion in powers of $q$, each of which is a polynomial in $T$. The Hilbert series and the Macdonald index are given in the {\tt Mathematica} notebook {\tt ArcSpaceHS.nb} attached to the arXiv submission of this paper. This provides strong evidence that the ring $S$ and the ideal $I_1$, in equations \eqref{eqn:A6S} and \eqref{eqn:A6I1}, respectively, capture the Schur sector of the $\widehat{E}_6(SU(2))$ SCFT as described in equation \eqref{eqn:biggereqn}.

\section{Discussion}\label{sec:disc}

In this paper, we have extended the proposal of \cite{Kang:2025zub,Kang:2026nge,Kang:2026ona,Kang:2026ney,Kang:2026jnv} to theories whose Schur sector contains fermionic operators that cannot be written in terms of other operators, by allowing the rings to be $\mathbb{Z}_2$-graded-commutative and, where necessary, the relations to involve derivatives of the generators. For all of the theories that we have studied except $\widehat{E}_6(SU(2))$, the ring is Zhu's $C_2$-algebra of the associated VOA, supplemented, for $\mathcal{N}=4$ super Yang--Mills with gauge algebra $\mathfrak{su}(2)$, by the stress tensor, and our results indicate that these VOAs are classically free; for $\mathfrak{su}(3)$, we have confirmed this directly up to and including conformal dimension five. For the rank-one $\mathcal{N}=3$ SCFTs and $\mathcal{N}=4$ super Yang--Mills, the ideal moreover follows from superpotentials, in analogy with the potentials of the Argyres--Douglas theories studied in \cite{Kang:2026nge,Kang:2026ona,Kang:2026ney,Kang:2026jnv}, and, in $\mathcal{N}=4$ superspace, from a single superfield. For $\widehat{E}_6(SU(2))$, classical freeness fails, and the Macdonald index is instead reproduced by the generalized arc space $J_{\infty,1}(S, I_1)$, in which three additional relations among the generators and their first derivatives are imposed.

These examples lead us to the following general expectation: for any 4d $\mathcal{N}=2$ SCFT, there exists a $\mathbb{Z}_2$-graded-commutative polynomial ring, $S$, together with an ideal, $I_k$, of the $k$th jet ring, $S_k$, for some $k \geq 0$, such that
\begin{equation}\label{eqn:evenbiggereqn}
	I_\text{Mac}(q, T) = \operatorname{HS}_{q,q,T,-1}(\operatorname{gr}(J_{\infty,k}(S, I_k))) \,,
\end{equation}
where the $T$-filtration is induced by assigning $T$-degrees to the generators of $S$ alone. Part of this expectation follows from the conjecture of Section \ref{sec:arcs}. Let the generators of $S$ correspond to the strong generators of the VOA, $\mathbb{V}$, and suppose, as conjectured there, that the kernel, $K$, of the map $S_\infty \rightarrow \operatorname{gr}_{\operatorname{Li}}(\mathbb{V})$ is finitely generated as a differential ideal, so that it is generated by relations of order at most $k$, for some $k$. If $I_k$ is the ideal of $S_k$ generated by these relations, then $J_{\infty,k}(S, I_k) = S_\infty \, / \, K \cong \operatorname{gr}_{\operatorname{Li}}(\mathbb{V})$, whose Hilbert series is the vacuum character, and so the Schur limit of equation \eqref{eqn:evenbiggereqn} holds. The non-trivial content of equation \eqref{eqn:evenbiggereqn} lies elsewhere. First, the Macdonald index is not determined by the VOA alone, and it is a strong constraint that it is reproduced by a $T$-filtration induced from the generators of $S$: for $\mathcal{A}(6)$, the $T$-degrees are fixed, up to the exchange of $\Psi$ and $\widetilde{\Psi}$, by a single coefficient of the Macdonald index, and equation \eqref{eqn:evenbiggereqn} then holds for the first fifty-two coefficients. Second, the order of the relations is small: for every SCFT that we have studied, $k \leq 1$ suffices, and, for $\mathcal{A}(6)$, we have conjectured that $K$ is generated by relations of order at most one. It would be interesting to determine whether there is a bound on $k$ that holds for all SCFTs, and to find a 4d $\mathcal{N}=2$ SCFT for which relations involving higher derivatives are necessary, as they are for the Ising model \cite{MR4589087}, or else to understand why they are not.

For $\mathcal{A}(6)$, the additional relations can be traced to normally ordered products, or combinations of them, such as $:\!\partial\Psi\Psi\!:$, which vanish identically in the free-field realization. Free-field realizations are known for the VOAs of large classes of 4d $\mathcal{N}=2$ SCFTs \cite{Bonetti:2018fqz}, and it would be interesting to determine whether such identically vanishing normally ordered products provide a general criterion for the failure of classical freeness, and whether they also determine the order, $k$, of the relations which must be imposed.

We now return to the $T$-filtration. For the $\mathcal{D}_p(G)$ Argyres--Douglas theories studied in \cite{Kang:2026ney}, the ring is the Jacobian ring of a single potential, and the $T$-filtration is intrinsic to this ring, given by the powers of the nilradical, so that the full filtered graded ring follows from the potential. For the theories in this paper, the $T$-degrees of the generators are instead fixed by the 4d charges of the corresponding Schur operators, and, for the rank-one $\mathcal{N}=3$ SCFTs and for $\mathcal{N}=4$ super Yang--Mills with gauge algebra $\mathfrak{su}(3)$, they are related by the supersymmetry variations $\delta_\pm$. As expected, the nilradical filtration, which is specific to the Argyres--Douglas theories, does not extend directly to the rings studied here, since, for example, the fermionic generators are all nilpotent but carry different $T$-degrees. It would be interesting to understand whether there is an analogue of the nilradical filtration for the $\mathbb{Z}_2$-graded-commutative rings of Sections \ref{sec:N4su3} and \ref{sec:N3}, whose ideals are generated by superpotentials, that is, whether the $T$-filtration can also be determined intrinsically from the superpotentials and their supersymmetry variations, so that, as for the Argyres--Douglas theories, the full filtered graded ring follows from the superpotentials. More generally, determining the $T$-filtration directly from the VOA, perhaps via the filtration associated to the 4d R-symmetry charges \cite{Song:2016yfd,Beem:2019tfp,ArabiArdehali:2025fad,Beem:2026lkq}, remains an important open problem. It would also be interesting to extend the analysis to $\mathcal{N}=4$ super Yang--Mills with gauge algebra $\mathfrak{su}(N)$ for $N > 3$, where one would need to determine which components of the higher $\mathcal{N}=4$ multiplets are strong generators, and to the higher-rank $\mathcal{N}=3$ SCFTs. In both cases, a superpotential construction, generalizing those of Sections \ref{sec:N4su3}, \ref{sec:N4superspace}, and \ref{sec:N3}, would provide a compact description of the ring; for $\mathfrak{su}(N)$, it would be interesting to determine the spin of the superspace superpotential, that is, the representation of the $\mathfrak{su}(2)$ flavor symmetry in which its lowest components transform.

The $\widehat{E}_6(SU(2))$ SCFT is the simplest example of a large class of 4d $\mathcal{N}=2$ SCFTs, $\widehat{\Gamma}(G)$, labelled by a pair of simply-laced Lie groups, which are obtained by gauging the diagonal $G$ flavor symmetry of a collection of $\mathcal{D}_p(G)$ Argyres--Douglas theories; conformal gaugings of this kind were studied early on in \cite{Cecotti:2012jx,Cecotti:2013lda}. For $\Gamma = D_4$, $E_6$, $E_7$, $E_8$ and particular choices of $G$, these theories have identical central charges, $a = c$ \cite{Kang:2021lic}; see, for example, \cite{Buican:2016arp,Buican:2020moo,Kang:2022vab,Jiang:2024baj,Pan:2025gzh} and references therein for further studies of these theories and of their chiral algebras. These theories generalize the $\mathcal{A}(6)$ example. Their Macdonald indices can be computed via the gauging formula, as in equation \eqref{eqn:refA6gauge}, from the Macdonald indices of the $\mathcal{D}_p(G)$ building blocks \cite{Song:2015wta}, and, when $\gcd(p, h^\vee) = 1$, these building blocks are among the Argyres--Douglas theories for which we have proposed filtered graded rings in \cite{Kang:2026ney}. The next example is $\widehat{E}_7(SU(3))$, which is obtained by gauging the diagonal $\mathfrak{su}(3)$ flavor symmetry of $\mathcal{D}_2(SU(3))$ and two copies of $\mathcal{D}_4(SU(3))$. Another family of interest is $\widehat{D}_4(SU(2n+1))$, obtained by gauging the diagonal $\mathfrak{su}(2n+1)$ flavor symmetry of four copies of $\mathcal{D}_2(SU(2n+1))$, whose Schur index can be written in terms of MacMahon's generalized sum-of-divisors function \cite{Kang:2021lic}, and which also admits a realization in class $\mathcal{S}$ of type $A_{2n}$, on a sphere with four $\mathbb{Z}_2$-twisted null punctures \cite{Beem:2020pry,Beratto:2020wmn,Kang:2022zsl}. We expect that the chiral algebras of the $\widehat{\Gamma}(G)$ theories share the features of $\mathcal{A}(6)$, including fermionic strong generators and, potentially, the failure of classical freeness, and thus that they will provide stringent tests of the generalized arc spaces $J_{\infty,k}(S, I_k)$.

The $a = c$ theories are also closely related to $\mathcal{N}=4$ super Yang--Mills: their Schur indices can be written in terms of those of $\mathcal{N}=4$ super Yang--Mills upon rescaling the fugacities \cite{Kang:2021lic}. For $\widehat{E}_6(SU(2))$ and $\mathcal{N}=4$ super Yang--Mills with gauge algebra $\mathfrak{su}(2)$, this rescaling reflects an exact graded vector space isomorphism between the associated VOAs, $\mathcal{A}(6)$ and the small $\mathcal{N}=4$ superconformal algebra with $c = -9$, which respects the $\mathfrak{u}(1)_r$ charge of the parent 4d theories \cite{Buican:2020moo}: the affine currents of the latter are mapped to the stress tensor of $\mathcal{A}(6)$ and some of its descendants, while the additional supercurrents are mapped to higher-dimensional fermionic currents and their descendants. Since we have determined the rings for both of these theories, and since one of the VOAs is classically free while the other is not, it would be interesting to understand whether this isomorphism has a counterpart at the level of the rings and their generalized arc spaces. There also exists an outer automorphism of the VOA of $\mathcal{N}=4$ super Yang--Mills, which allows us to consider a twisted character that turns out to be related to non-invertible symmetries \cite{Maruyoshi:2026cmr}. It would be interesting to understand how this automorphism is reflected in the $a=c$ theories as well. More generally, it would be desirable to understand how the ring associated to a gauge theory, such as $\widehat{E}_6(SU(2))$ or $\mathcal{N}=4$ super Yang--Mills, arises from the rings associated to its constituents. For example, there exists a commutative filtered graded ring which reproduces the Macdonald index of $\mathcal{D}_3(SU(2))$ via equation \eqref{eqn:bigeqn} \cite{Andrews:2025krn,Kang:2025zub}, and it would be interesting to understand whether the ring of $\widehat{E}_6(SU(2))$ can be obtained from three copies of it via a ``gauging'' procedure that mirrors equation \eqref{eqn:refA6gauge}.

\vspace{0.3cm}
\subsection*{Acknowledgements}
M.~J.~K.~is supported by the DOE through DE-SC0010813 and the Start-up Research Grant for new faculty provided by Texas A\&M University.
C.~L.~acknowledges the Deutsche Forschungsgemeinschaft under Germany's Excellence Strategy - EXC 2121 ``Quantum Universe'' - 390833306 and the Collaborative Research Center - SFB 1624 ``Higher Structures, Moduli Spaces, and Integrability'' - 506632645. The work of H.~H.~and J.~S.~is supported by the National Research Foundation of Korea (NRF) grants RS-2024-00405629 and RS-2026-25482546, and the KAIST-KIAS collaboration program. The work of J.~S.~is also supported in part by the Walter Burke Institute for Theoretical Physics and the Leinweber Forums for Theoretical Physics at Caltech and by the U.S.~Department of Energy, Office of Science, Office of High Energy Physics, under Award Number DE-SC0011632.

\bibliography{references}{}
\bibliographystyle{sortedbutpretty}

\end{document}